\documentclass[twocolumn]{raa}           
\usepackage{graphicx,times}
\usepackage{natbib}
\usepackage{amssymb,amsmath}
\bibpunct{(}{)}{;}{a}{}{,}

\usepackage[a4paper=true,dvipdfm=true,pagebackref=true]{hyperref}
\hypersetup{pdftitle = The title of my PDF, pdfauthor = My name, pdfsubject= The subject, pdfkeywords = keyword1 keyword2 keyword3} 
\hypersetup{colorlinks = true, linkcolor = green, anchorcolor = red, citecolor = blue, filecolor = red, pagecolor = red, urlcolor = red}

\begin{document}

   \title{ The Distance Measurements of  Supernova Remnants in The Fourth Galactic Quadrant 
$^*$
\footnotetext{\small $*$ Supported by the National Natural Science Foundation of China.}
}

 \volnopage{ {\bf 20XX} Vol.\ {\bf X} No. {\bf XX}, 000--000}
   \setcounter{page}{1}

   \author{Su-Su Shan\inst{1,2}, Hui Zhu \inst{1,3}, Wen-Wu Tian\inst{1,2}, Hai-Yan Zhang\inst{1}, Ai-Yuan Yang\inst{4}, Meng-Fei Zhang\inst{1,2}}

   \institute{ National Astronomical Observatories, Chinese Academy of Sciences, 20A
Datun Road, Chaoyang District, Beijing 100012, China; {\it shansusu@nao.cas.cn; \it zhuhui@nao.cas.cn}\\
        \and
        School of Astronomy, University of Chinese Academy of Sciences,
Beijing 100049, China \\
       \and
       Harvard-Smithsonian Center for Astrophysics, 60 Garden Street, Cambridge, MA 02138, USA \\ 
       \and
       Max-Planck-Institut f\"ur Radioastronomie, Aufdem H\"ugel 60, D-53121, Bonn, Germany\\
     \vs \no
   {\small Received 2018 ; accepted 2019}
}

\abstract{ We take advantage of the red clump stars to build the  relation of the optical extinction ($\rm A_V$) and distance in each direction of supernova remnants (SNRs) with known extinction in the fourth Galactic quadrant. The distances of 9 SNRs  are well determined by this method. Their uncertainties range from 10\% to 30\%, which is significantly improved for 8 SNRs, G279.0+1.1, G284.3-1.8, G296.1-0.5, G299.2-2.9, G308.4-1.4, G309.2-0.6, G309.8-2.6, G332.4-0.4. In addition, SNR G284.3-1.8 with the new distance of 5.5 kpc is not likely associated with the PSR J1016-5857 at 3 kpc.
}

   \authorrunning{S.-S. Shan et al. }            
   \titlerunning{Distances of SNRs}  
   \maketitle

%
\section{Introduction}           
\label{sect:intro}
Distance is a basic and important parameter for a supernova remnant (SNR) to constrain its size, age, expansion velocity and the explosion energy of the progenitor supernova \citep[e.g.,][]{Tian2008,Zhou2018}.
However, it is challenging to obtain reliable distances of  supernova remnants. About 30\% of SNRs' distances are available in the Galactic SNRs' catalog of \cite{Green2014a} and $\sim$ 50\% of SNRs have distances in the catalogue of \cite{Ferrand2012}. Most of these distances are not given the uncertainty estimates. 

There are several methods to measure the distance of SNRs. Firstly, the kinematic distances of SNRs are likely to be measured based on their HI absorptions or molecular line emissions from the clouds interacting with them \citep[e.g.,][]{Leahy2010,Su2011,Ranasinghe2018}. Secondly, for shell-type radio SNRs, distances can be estimated by the $\Sigma$-D relation \citep[e.g.,][]{Case1998}. Thirdly, the distances of SNRs are usually equivalent to the distances of their possible associations, for examples, OB associations \citep[e.g.][Vela remnant]{Cha1999}, pulsars \citep[e.g.][]{Cordes2002}. For some rare SNRs, distances can also be obtained by proper motion \citep[e.g.][Kepler's supernova remnant]{Vink2008,Katsuda2008}, Sedov estimates \citep{Bocchino2000}, blast wave method \citep{McKee1975}, or extinction measurement \citep{Chen2017,Zhao2018}.

Red clump (RC) stars are characterized  by  concentrating in an obvious region of the colour-magnitude diagrams (CMDs) \citep{Gao2009}. They are usually low mass stars in the early stage of core-He burning and widely used to trace distances, and probe extinction towards Galactic objects since the dispersion of  their absolute magnitude and the intrinsic colour are small \citep{Girardi2016}. The relation of extinction and distance has been  used to estimate the distances of the objects with known extinction, such as pulsars \citep{Durant2006}, Low-mass X-ray binaries \citep{Tolga2010}. \citet{Zhu2015} first applied the RC method to determine the distance of SNR G332.5-5.6. \cite{Shan2018} applied the similar method to systematically analyse the extinction distances towards SNRs in the first Galactic quadrant and  independently obtained new extinction distances of 15 SNRs.  Their results have proved that the distances traced  by RC stars are reliable and consistent  with the kinematic distances. This paper presents our new results aiming at the SNRs in the rest of Galactic quadrants.

\section{Method}

\begin{figure}
\centering
\begin{tabular}{cc}
\includegraphics[angle=0,width=0.5\textwidth]{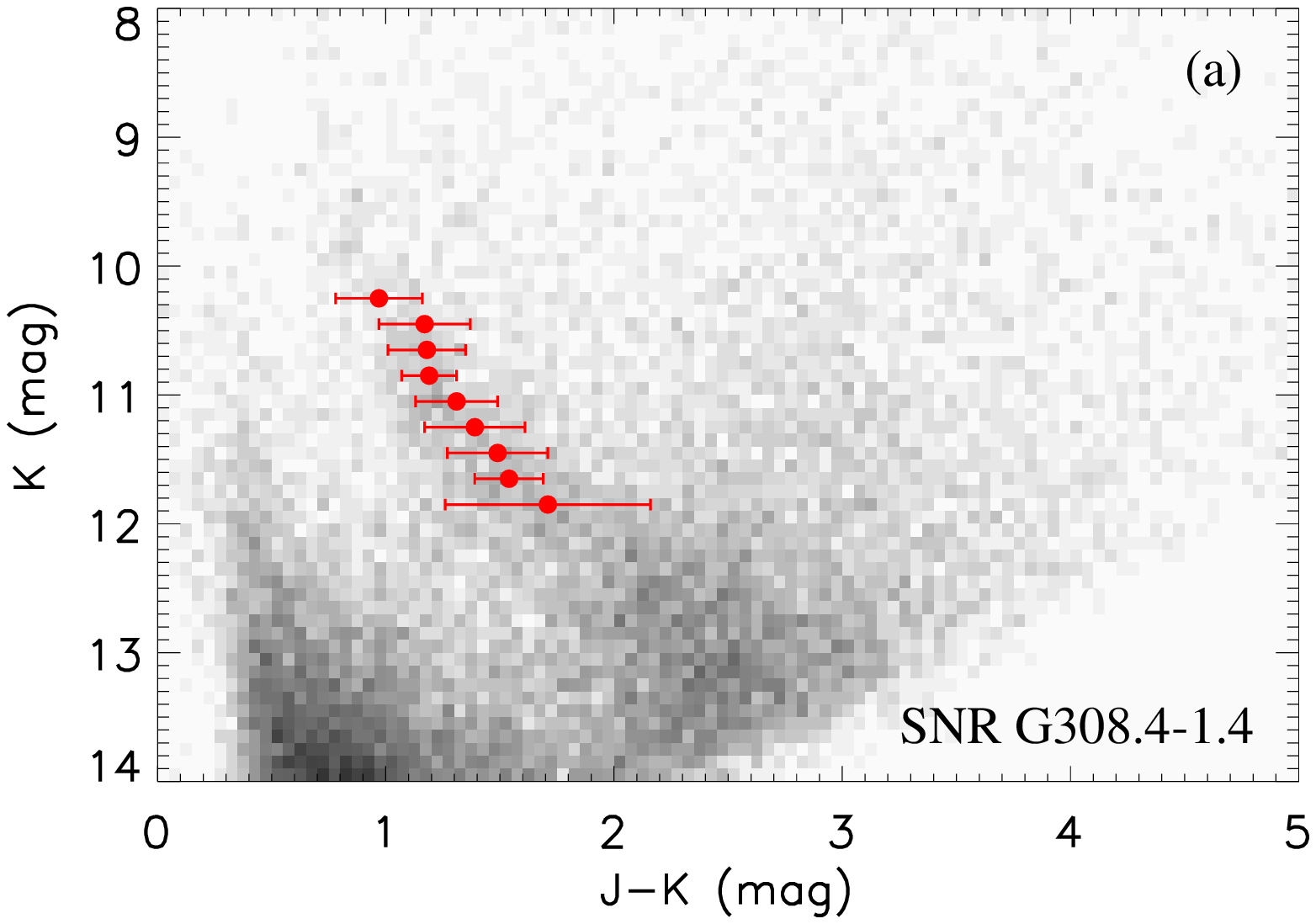}
\includegraphics[angle=0,width=0.5\textwidth]{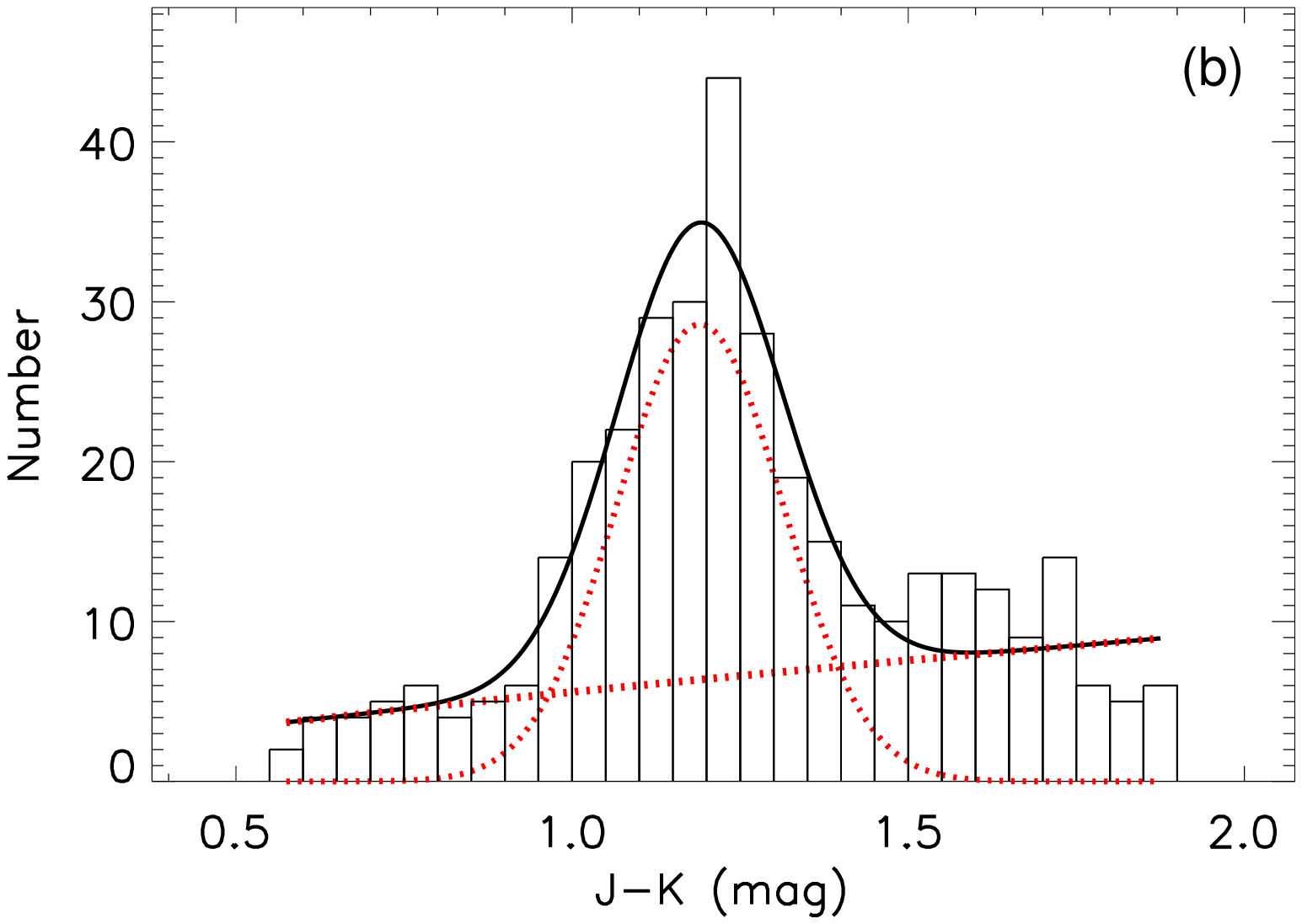}\\
\includegraphics[angle=0,width=0.5\textwidth]{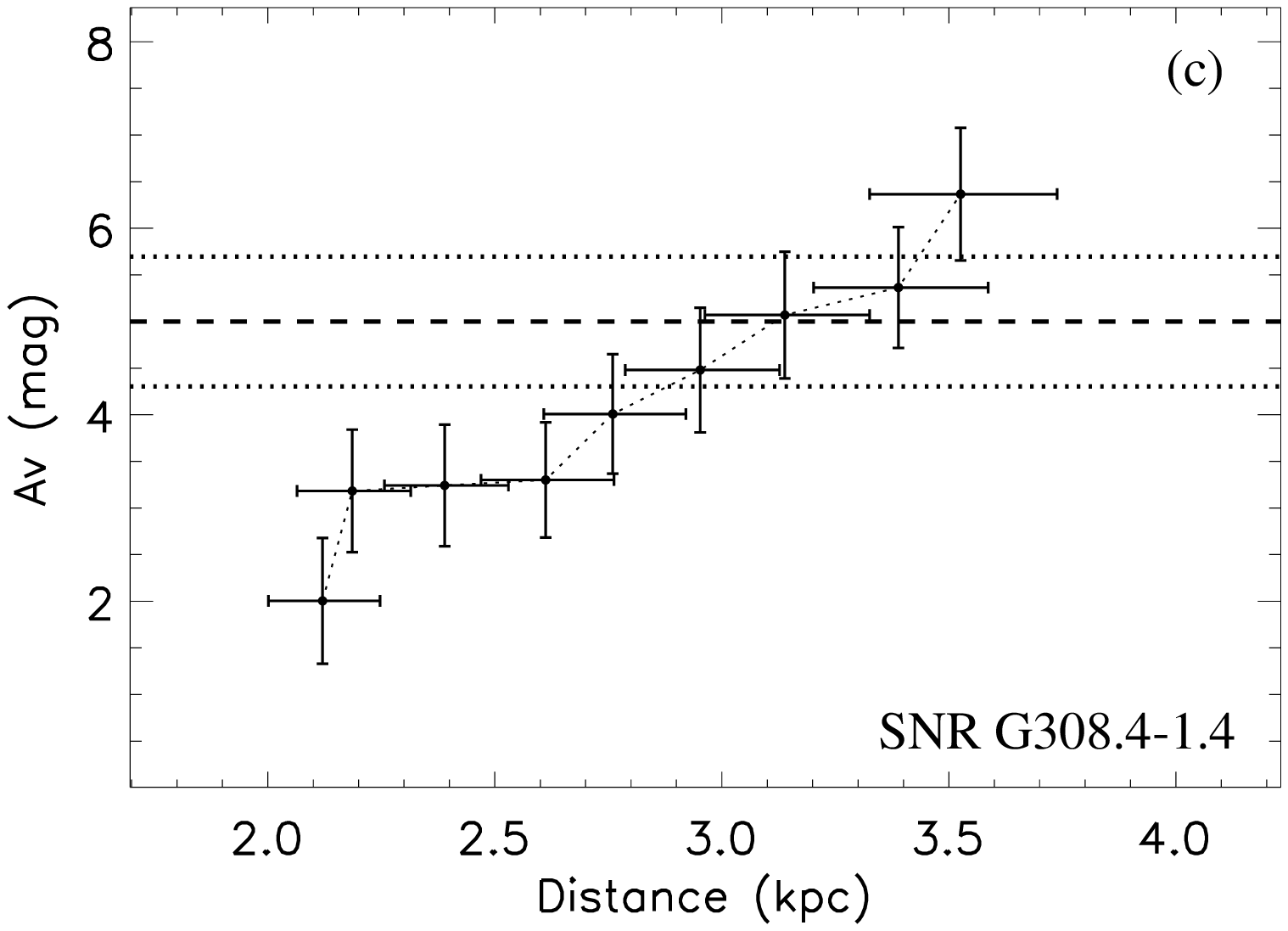}
\includegraphics[angle=0,width=0.5\textwidth]{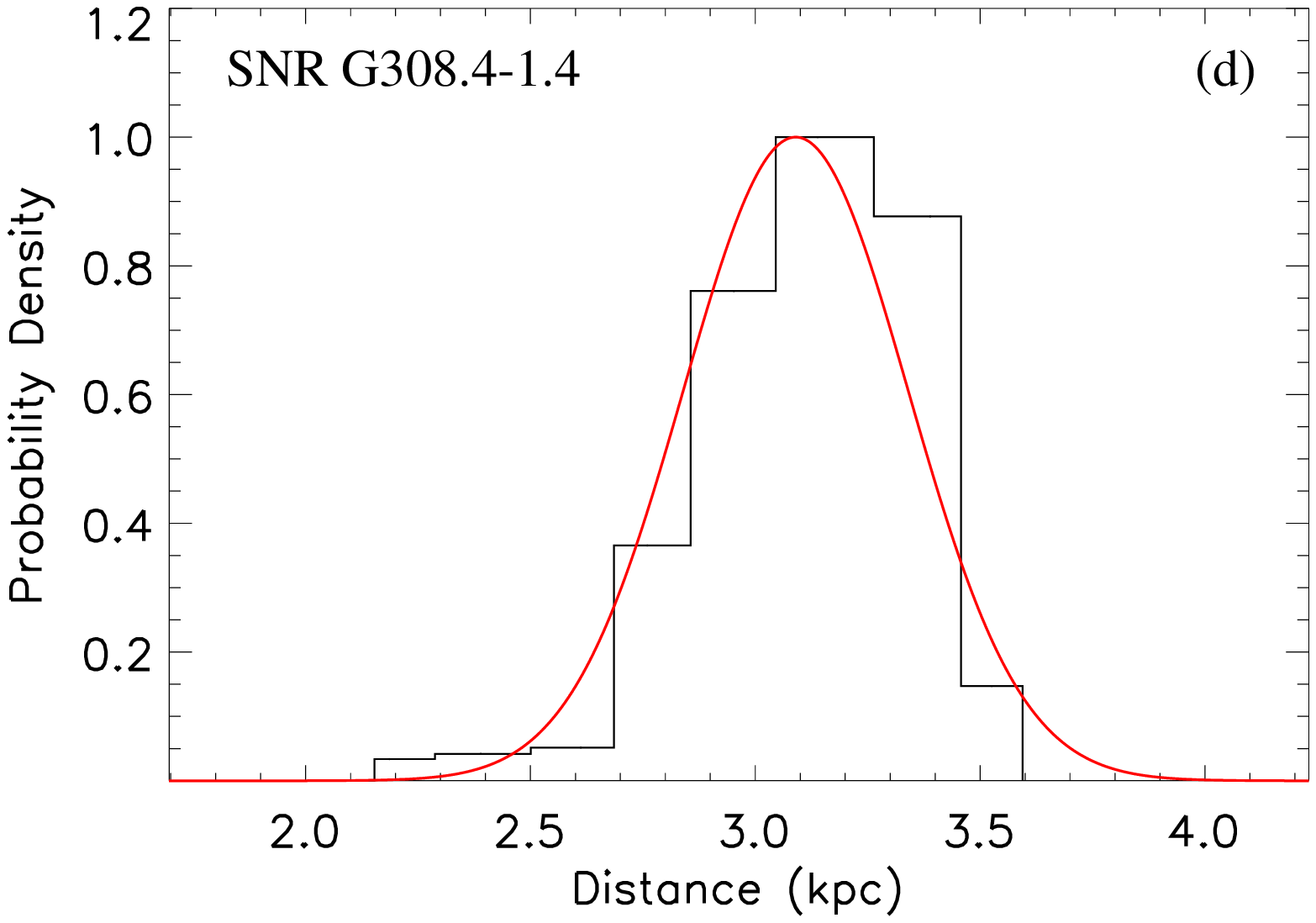}
\end{tabular}
\caption{(a) The CMD in the direction of G308.4-1.4 within 0.5 deg$^2$,
the red dot  and  lines show the fitted location of the RC peak density and its extent with $1\sigma$. (b) Histogram of the $\rm J-K$ values of the selected stars
 in the $\rm 10.6 <K < 11.1$. The black curve is the best fit to this histogram. The red dotted curves are the Gaussian and power-law components, respectively. (c) The $\rm A_V$-D relation in the direction of G308.4-1.4. The dashed line is $\rm A_v$ value of this SNR. The dotted lines are the uncertainties of $\rm A_v$. (d) Probability distribution over distance to the SNRs and the best-fit Gaussian model.}
\label{fig1}
\end{figure}

 We follow the RC method of \citet{Shan2018}. Here we briefly describe this method.
  To select the RC stars,  we build the Ks (hereafter K) vs. J-K CMD  in the direction of each  SNR with the
 2MASS All-Sky Point Source Catalogue \citep{Skrutskie2006}. The area around the SNR is $ 1^{\circ}$
$\times$  $ 0.5^{\circ}$ ($\bigtriangleup l\,$ $\times$ $\bigtriangleup b \,$). Fig. \ref{fig1} (a) shows an example of SNR 308.2-1.4. 
 
 We bin the stars sample  into a number of horizontal strips according to K magnitude. The sample of each bin is constructed as the stars count histogram. Then we use the function below to fit the histogram
\citep{Durant2006}:
\begin{equation}
\rm N=A_{RCs}exp\{ \frac{-[(J-K)-(J-K)_{peak}]^2}{2\sigma ^2}\}+A_{C}(J-K)^{\alpha}
\end{equation}
Here ($\rm J-K$) represents the stellar colour, $\rm A_{RCs}$, and $\rm A_{C}$ stand for
 the normalizations of RC stars and contaminant stars, respectively.
 The first term of this function is a Gaussian function to fit the  RC stars distribution;
the second term is a power law to fit the contaminant stars (see Fig.~\ref{fig1} (b)). 
The best fit yields the stellar colour of RC stars at peak density $(J-K)_{\rm peak}$ for each strip. The extinction  and distance are calculated from 
\begin{equation}
\rm A_K=0.67 \times[(J-K)_{peak}-(J-K)_{0}]
\end{equation}
\begin{equation}
\label{f2}
\rm \frac{A_K}{A_V}=0.1615-\frac{0.1483}{R_V}
\end{equation}
\begin{equation}
\rm D(kpc)=10^{[0.2(m_K-M_K+5-0.1137A_V)]}/1000
\end{equation}
 Here A$_{\rm V}$ and $\rm A_K$ are extinction in V and K bands.
 We adopt the total to selective extinction ratio $\rm R_V=3.1\pm0.18$ \citep{Schlafly2016}, the intrinsic colour of RC stars $\rm (J-K)_{0}=0.63\pm0.1$ mag \citep[e.g.,][]{Yaz2013,Grocholski2002} and the absolute magnitude of RC stars $\rm M_K=-1.61\pm0.1$  mag \citep[e.g.,][]{Alves2000,Grocholski2002,Hawkins2017}. The uncertainties of $R_V$, $(J-K)_{0}$ and $M_K$ are transferred to $A_V$ and D with error propagation formula (see \citet{Shan2018} and the references herein).
 
 In Figure\ref{fig1} (c), the relation between the optical extinction $\rm A_V$ and distances D (hereafter, $\rm A_V$-D)  is built along the direction of G308.4-1.4. Combining the $\rm A_V$ value of G308.4-1.4, its distance is estimated by  the Bayesian method \citep{{Tolga2010}}.
Fig.~\ref{fig1} (d) shows the probability distribution over distance calculated by  \begin{eqnarray}\label{pp}
P(D)= \int P_{SNR}(A_{K})P_{RC}(D|A_{V})dA_{V}.
\end{eqnarray}
Here $\rm P_{SNR}(A_{V})$ presents the probability distribution of SNR's extinction.  We assume $\rm P_{RC}(D|A_{V})=P_{RC}(A_{V}|D)$.  $\rm P_{RC}(A_{V}|D)$ presents the distribution of the extinction traced by RC  at each distance bin. Both distributions are denoted as Gaussian functions. Then we fit these distributions by Gaussian function, yielding the distance with the highest probability. For the objects with good Gaussian fitting, the uncertainty of distance is equal to the standard deviation of the Gaussian. However, for some objects, there are apparent and sudden decreases in the distance probability. The red lines mark such decreases (see Fig. 3 ). In this case, the uncertainty of distance reflects the cut-off distance.

\section{results and brief discussion}
\label{sect:Obs}
New  distances of 9 SNRs are obtained. The results are summarized in Table \ref{tab1} and  \ref{tab2}. We also note  that this method does not work for SNRs in the 2nd and 3rd quadrants after tens of trials. It might be caused by the fact that the extinction towards our targets  in the two  quadrants grows slowly with the increasing distance, which makes the $\rm A_V$-D relation too flat to give a reliable distance. Hence, we just discuss the results of the 4th quadrant. Fig. 2   presents the CMDs with the locations of RC's peak density in the direction of SNRs. Fig. 3   shows the corresponding A$_{\rm V}$-D relations and probability distribution over distance to the SNRs and the best-fit Gaussian model. We discuss them in detail below.

\begin{table}
\bc
\begin{minipage}[]{0.95\textwidth}
\caption[]{Optical extinction $\rm A_V$ and distances \label{tab1}}\end{minipage}
\setlength{\tabcolsep}{8pt}
\small
 \begin{tabular}{clllll}
  \hline\noalign{\smallskip}
 Source     &              $\rm A_V$   &       Distances   &  Previous measurements &  Method &           Ref.\\
  Name             &               mag  &              kpc           &    kpc        &               &          \\
  \hline\noalign{\smallskip}
   G279.0+1.1   & 1.6$\pm0.1$  &     2.7$\pm0.3$    &  3 & Blast wave \& $\Sigma$-D               &1\\
 G284.3-1.8    & 4.4$^1$  &     5.5$\pm0.7$    &   3, 5.6$_{-2.1}^{+4.6}$& distance of association&2, 3, 4  \\
G296.1-0.5    &     1.9$\pm$0.3    &      4.3$\pm0.8$   &  3$\pm$1.0   & reddening measurement  & 5, 6   \\
	      &                    &                            &  3.8$\pm$1.9 & kinematic measurement  & 5, 6  \\		   
	      &                    &                            &   7.7, 6.6, 4.9& $\Sigma$-D              & 7, 8, 9  \\			     
G315.4-2.3    &     1.7$\pm$0.2    &     $\leq 2.0$    &    2.8$\pm0.4$, 2.3$\pm0.2$ &kinematic measurement&10, 11, 12\\
              &                    &                            & $1.2\pm0.2$  &  Sedov estimate         &13\\
G332.4-0.4    &     4.7$\pm$0.7    &       3.0$\pm0.3$  &    3.3, 3.1-4.6       &  kinematic measurement  &14, 10, 15   \\
              &                    &                            &     6.5      &   extinction measurment  &16     \\		   
  \noalign{\smallskip}\hline
\end{tabular}
\ec
\tablecomments{0.98\textwidth}{ $^1$ We assume  the error of $A_V$ as 10\%  when determining the distance of G284.3-1.8.   }
\tablerefs{0.98\textwidth}{(1) \citet{Stupar2009}; (2)\citet{Abramowski2012}; (3) \citet{Kargaltsev2013};  (4)\citet{Napoli2011}; (5) \citet{Castro2011}; (6) \citet{Longmore1977}; (7)  \citet{Caswell1983}; (8) \citet{Case1998}; (9) \citet{Clark1973}; (10) \citet{Zhu2017};
(11) \citet{Rosado1996}; (12) \citet{Sollerman2003}; (13) \citet{Bocchino2000};  (14) \citet{Caswell1975}; (15) \citet{Reynoso2004}; 
(16) \citet{Ruiz1983}.}
\end{table}
\begin{table}
\bc
\begin{minipage}[]{100mm}
\caption[]{$\rm A_V$ converted from hydrogen column density $\rm N_H$ and distances \label{tab2}}\end{minipage}
\setlength{\tabcolsep}{5pt}
\small
 \begin{tabular}{cccllll}
  \hline\noalign{\smallskip}
 Source     &               $\rm N_H$   & $\rm  A_V^{1}$&      Distances  & Previous measurements &  Method &           Ref.\\
  Name            &           10$\rm ^{21}Hcm^{-2}$ & mag&             kpc           &    kpc        &               &          \\
  \hline\noalign{\smallskip}
G299.2-2.9    &     3.5$\pm$1.5  &1.7$\pm0.7$    &     2.8$\pm0.8$    &   0.5, 2-11   &   hydrogen column density &1, 2  \\
G308.4-1.4    &    10.2$\pm$1.4  &5$\pm0.7$      &     3.1$\pm0.3$    &   5.9$\pm2.0$&    extinction estimate    & 3\\
              &                    &             &              &   $2.0\pm0.6$, 12.5$\pm$0.7& kinematic measurement  &3\\
              &                    &             &               &   9.8$^{+0.9}_{-0.7}$& Sedov estimate&3\\
G309.2-0.6    &     6.5$\pm$3.0  &3.2$\pm1.5$    &      2.8$\pm0.8$   &   $4\pm2$    &    hydrogen column density &4\\
              &                    &             &              &  5.4$\pm1.6$ to 14.1$\pm0.7$ & kinematic measurement&5\\
G309.8-2.6    &     3.9$\pm$0.4   &1.9$\pm0.2$ &      2.3$\pm0.2$   &  2.5         &    pulsar distance & 6, 7\\
  \noalign{\smallskip}\hline
\end{tabular}
\ec
\tablecomments{0.95\textwidth}{$^1$ $\rm A_V$ =$\rm N_H$/(2.04$\pm0.05$)$\times$ 10$^{21}$cm$^{-2}$mag$^{-1}$ \citep{Zhu2017}.}
\tablerefs{0.98\textwidth}{(1) \citet{Park2007}; (2) \citet{Slane1996};
(3) \citet{Prinz2012}; (4) \citet{Rakowski2001}; (5) \citet{Gaensler1998}; (6) \citet{Lemoine-Goumard2011}; (7) \citet{Camilo2004}.}
\end{table}

\begin{figure}

\caption{The CMDs within 0.5 deg$^2$ in  the  directions of 8 SNRs, the grey colours denote stellar densities in the logarithmic scale. The red dots and  lines show the fitted location of the RC peak density and its extent with $1 \sigma$.}
\begin{tabular}{cc}
\includegraphics[width = 0.5\textwidth]{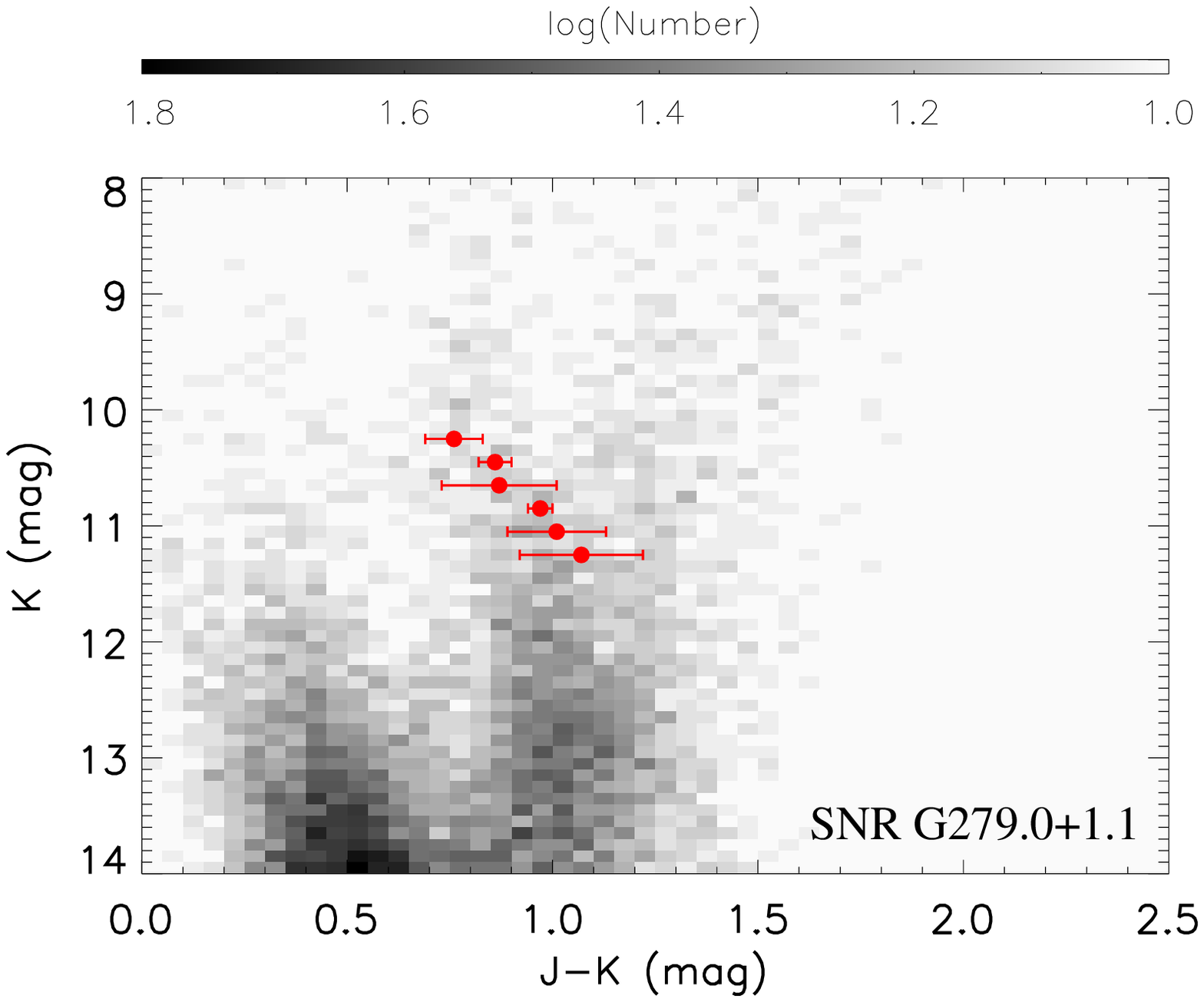}&
\vspace*{0.3cm}
\includegraphics[width = 0.5\textwidth]{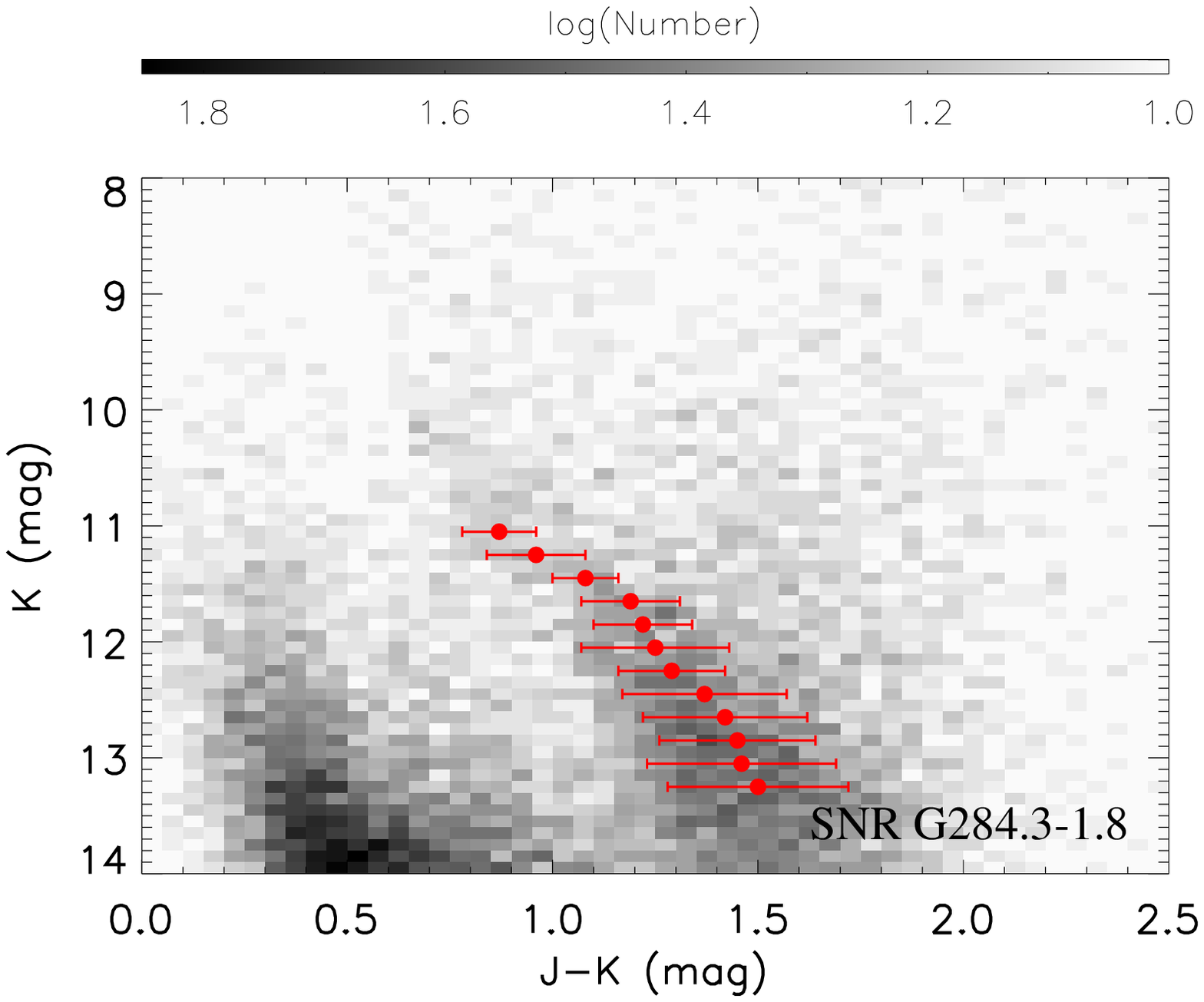}\\
\vspace*{0.3cm}
\includegraphics[width = 0.5\textwidth]{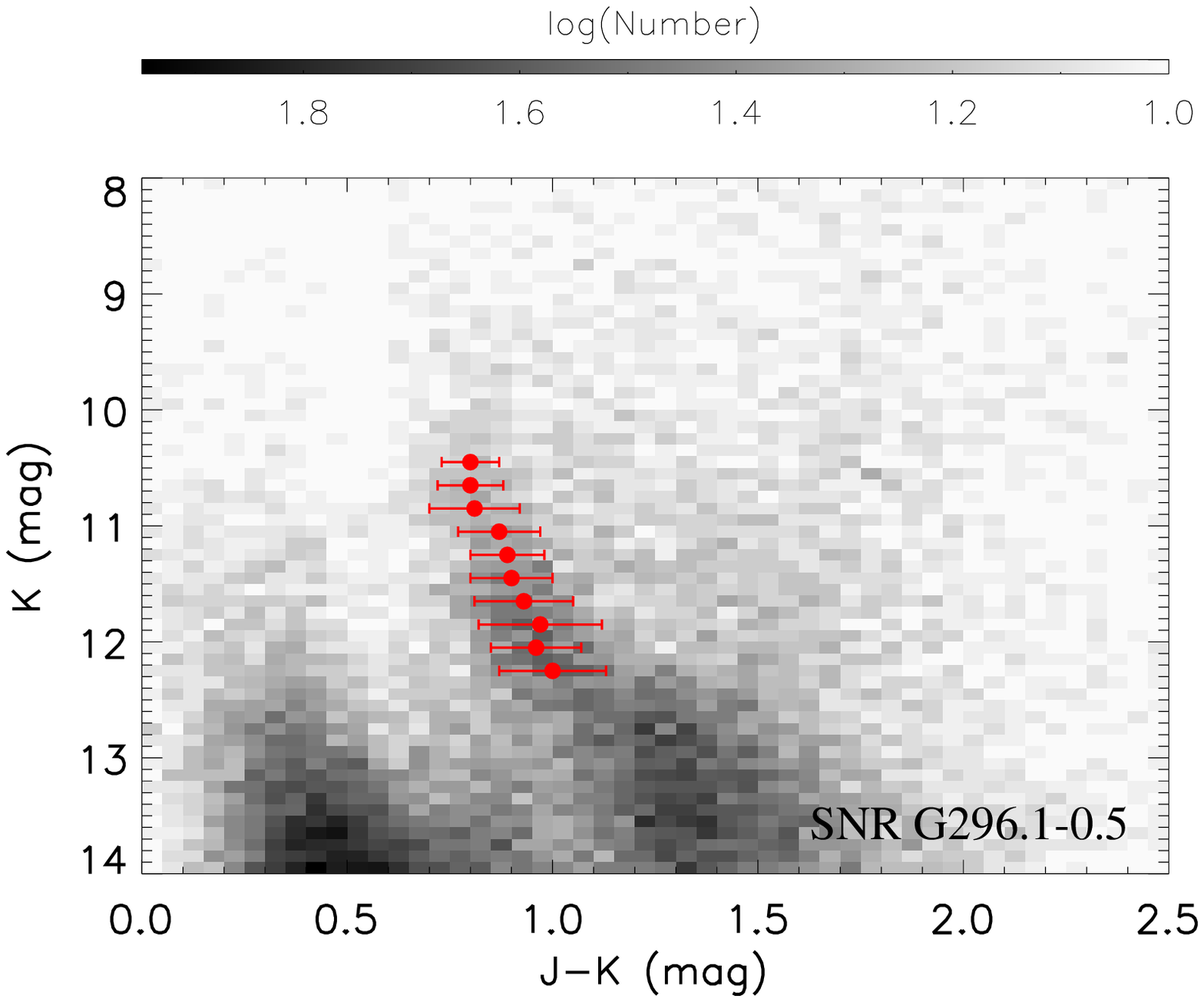}&
\vspace*{0.3cm}
\includegraphics[width = 0.5\textwidth]{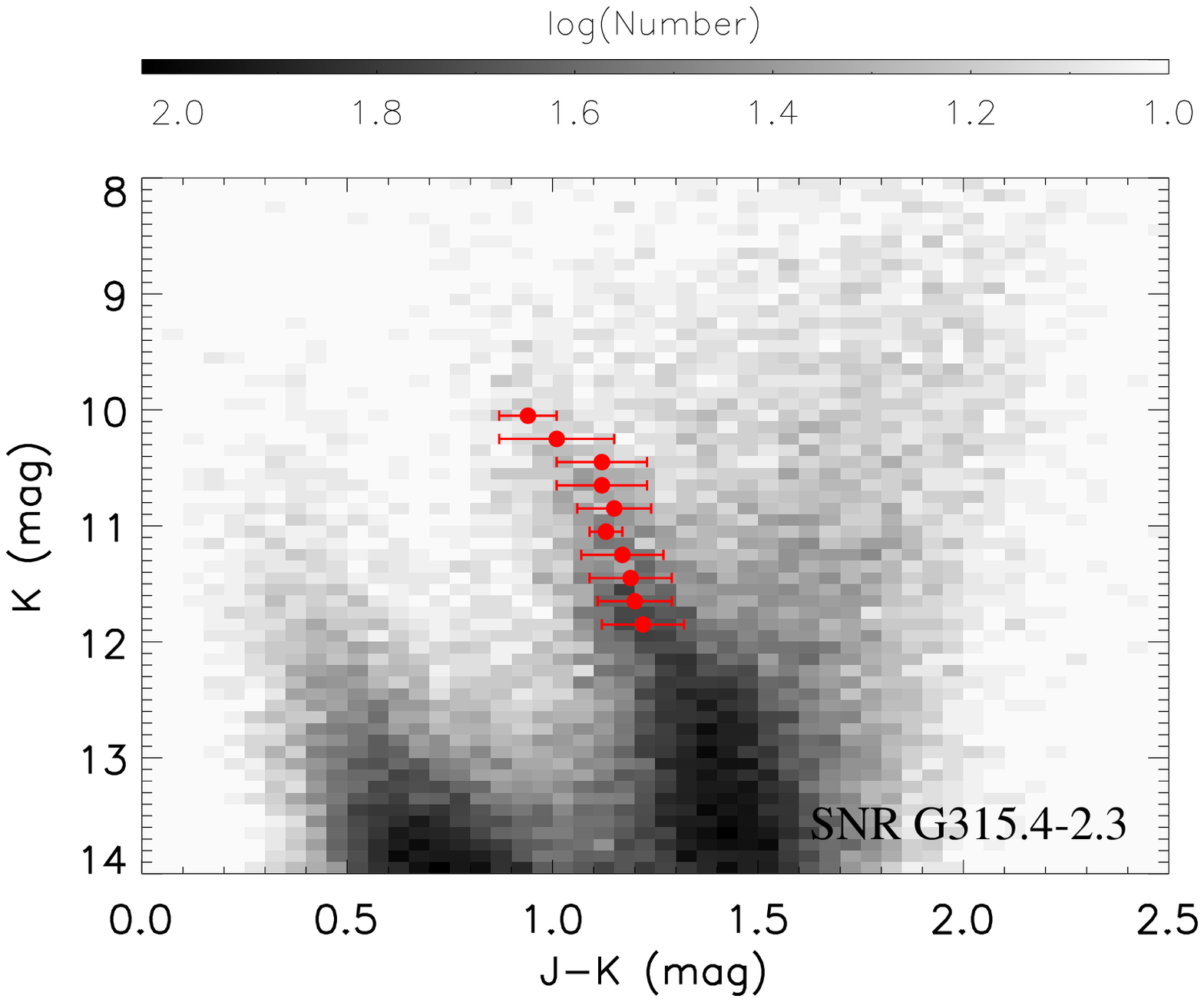}\\
\vspace*{0.3cm}
\includegraphics[width = 0.5\textwidth]{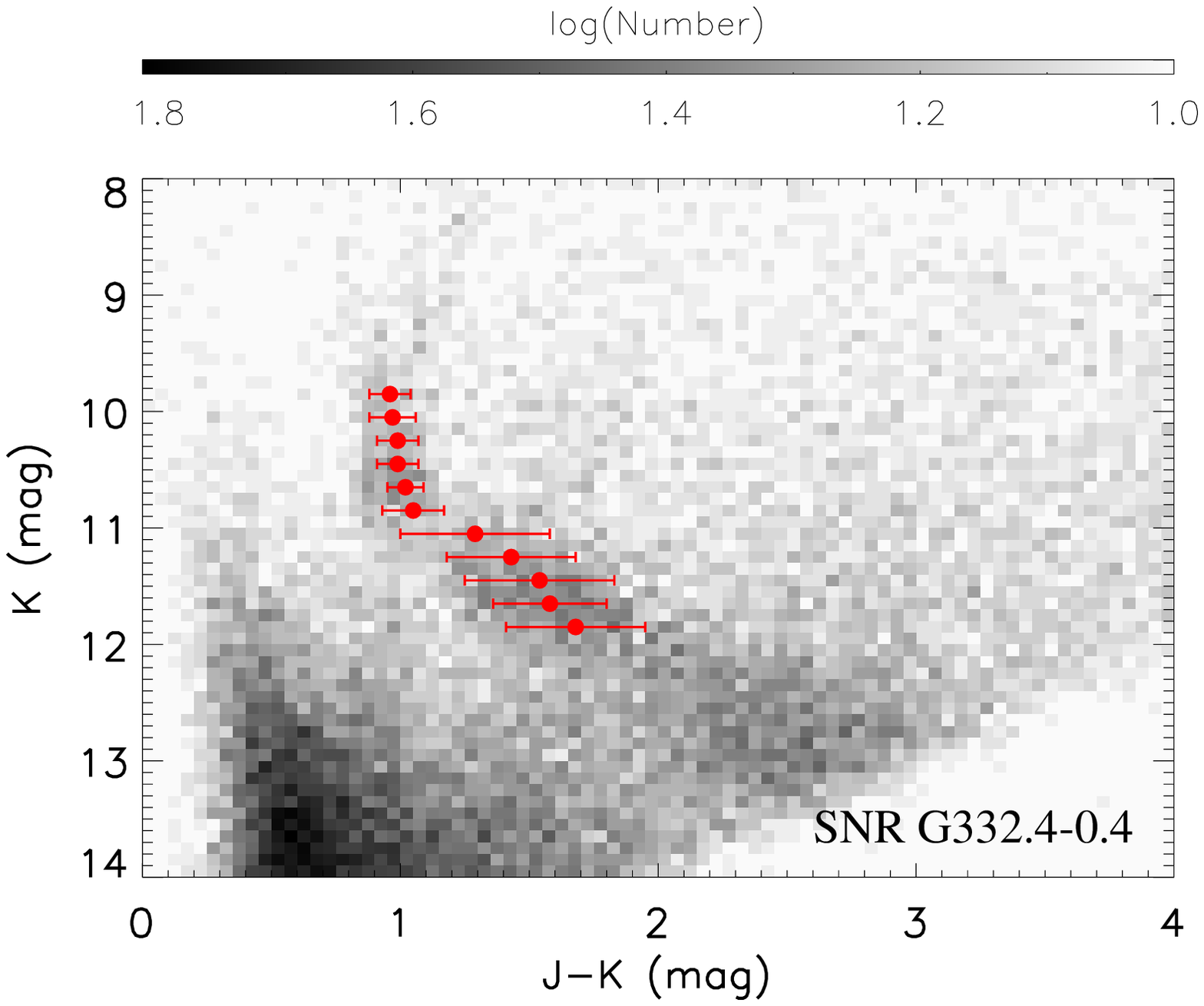}&
\vspace*{0.3cm}
\includegraphics[width = 0.5\textwidth]{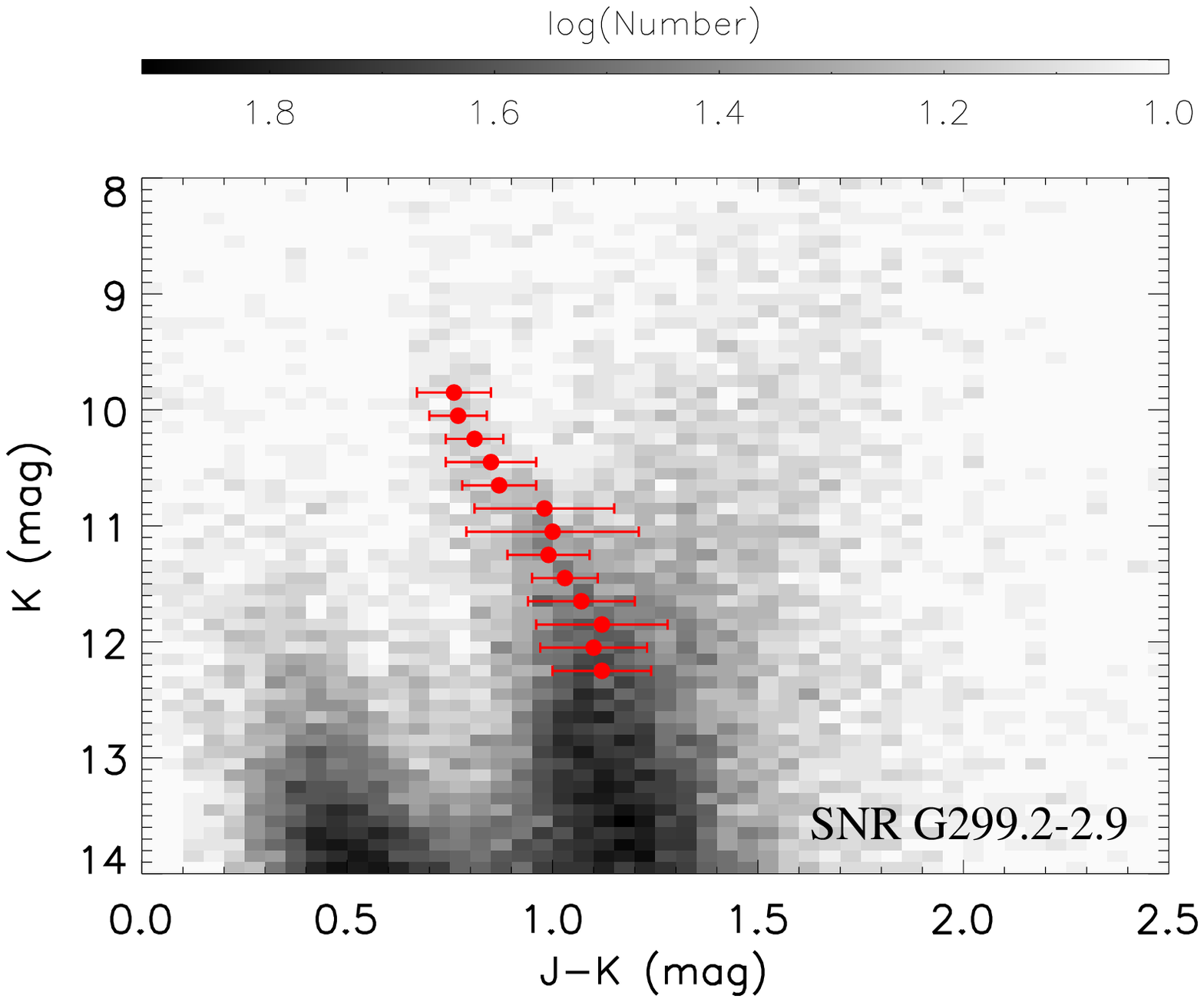}\\
\vspace*{0.3cm}
\includegraphics[width = 0.5\textwidth]{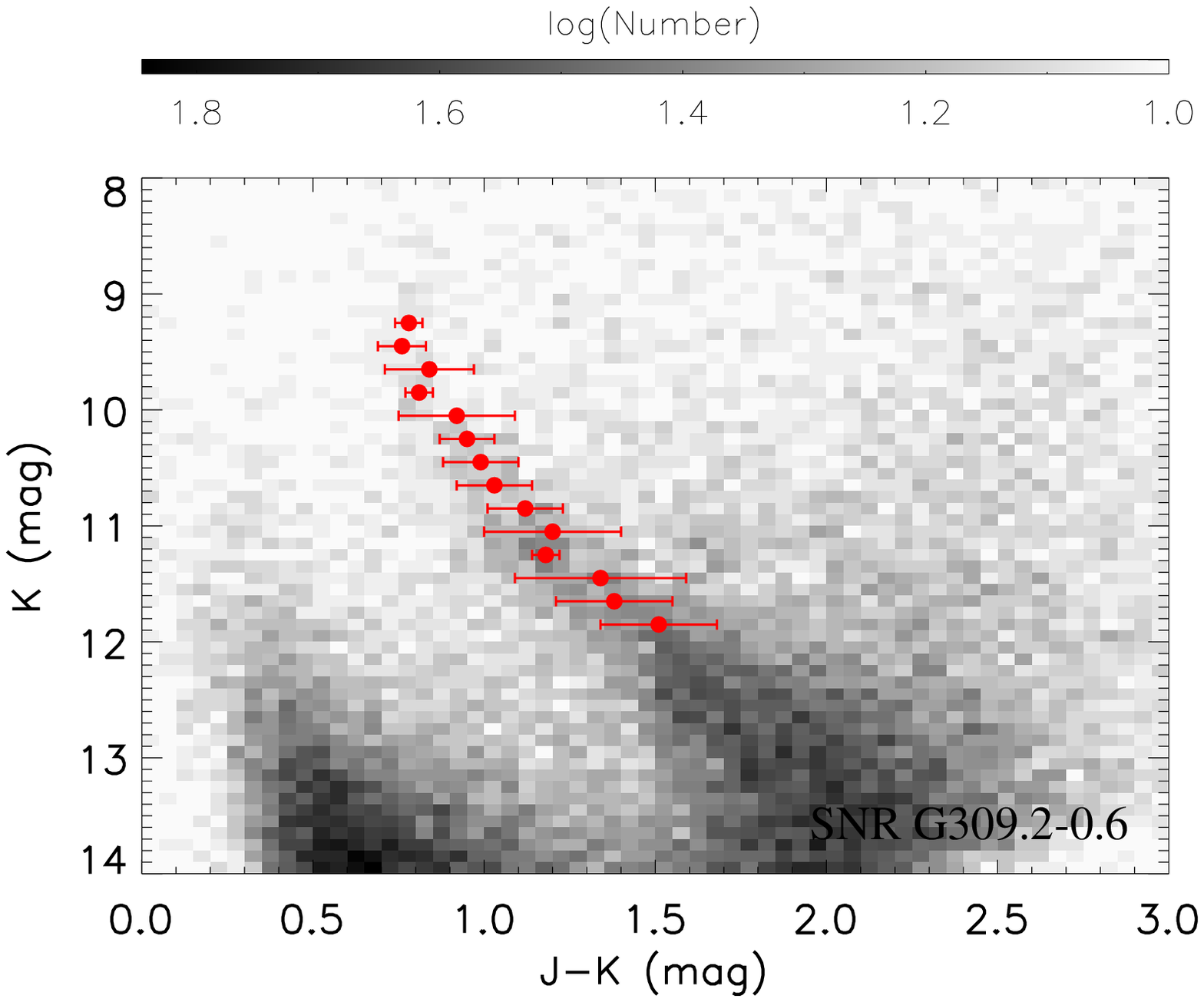}&
\vspace*{0.3cm}
\includegraphics[width = 0.5\textwidth]{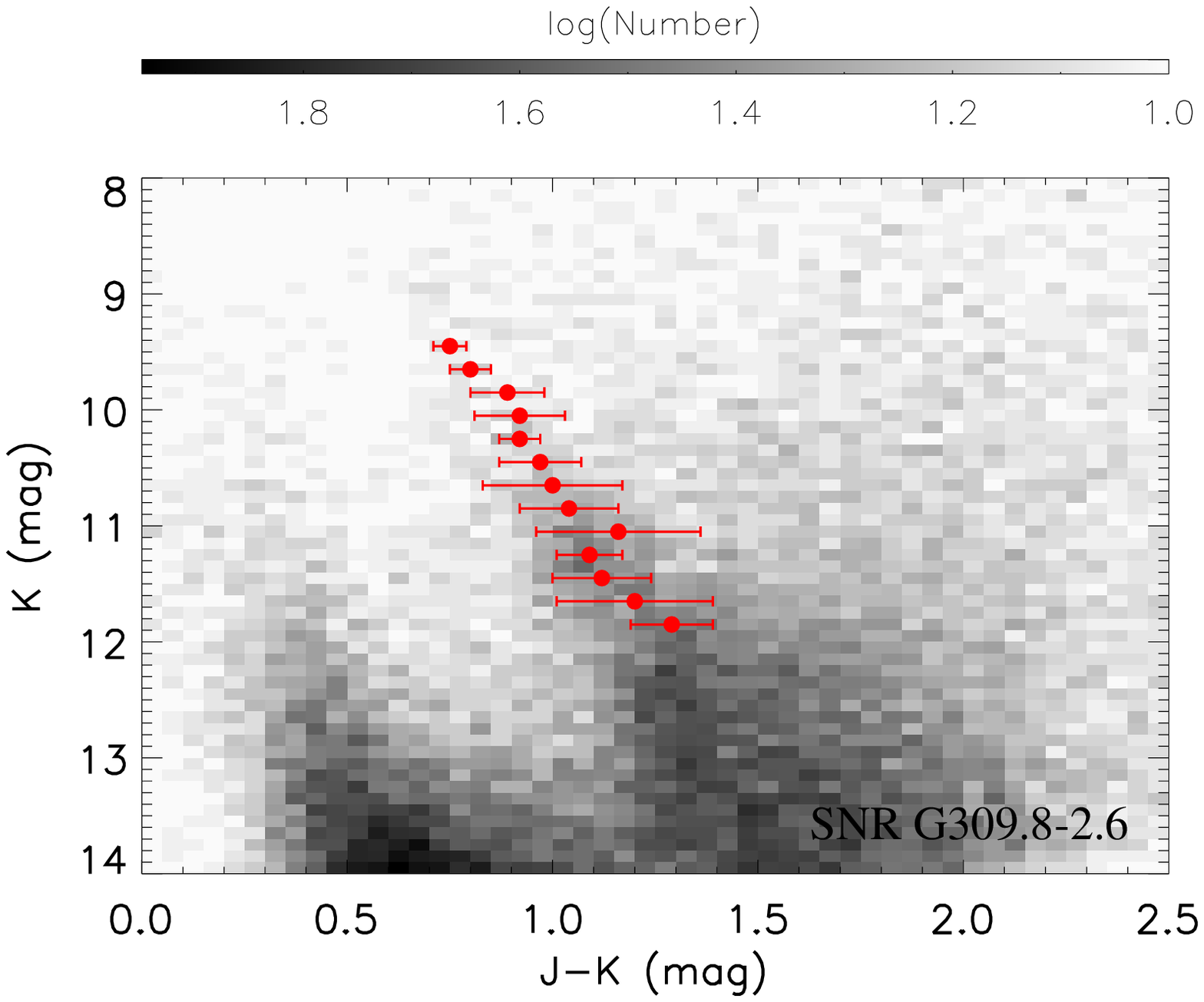}
\end{tabular}
\label{fig2}
\end{figure}

\begin{figure}
\caption{Left: The $\rm A_V$-D relation  traced by RC stars along the direction of each SNR. The dashed line is $\rm A_V$ value of each SNR. The dotted lines are the uncertainties of $\rm A_V$. Right: Probability distribution over distance to the SNRs and the best-fit Gaussian model with the cutoffs. The red line in the right panel for G315.4-2.3 is the upper distance limit. }
\begin{tabular}{cc}
\includegraphics[width = 0.5\textwidth]{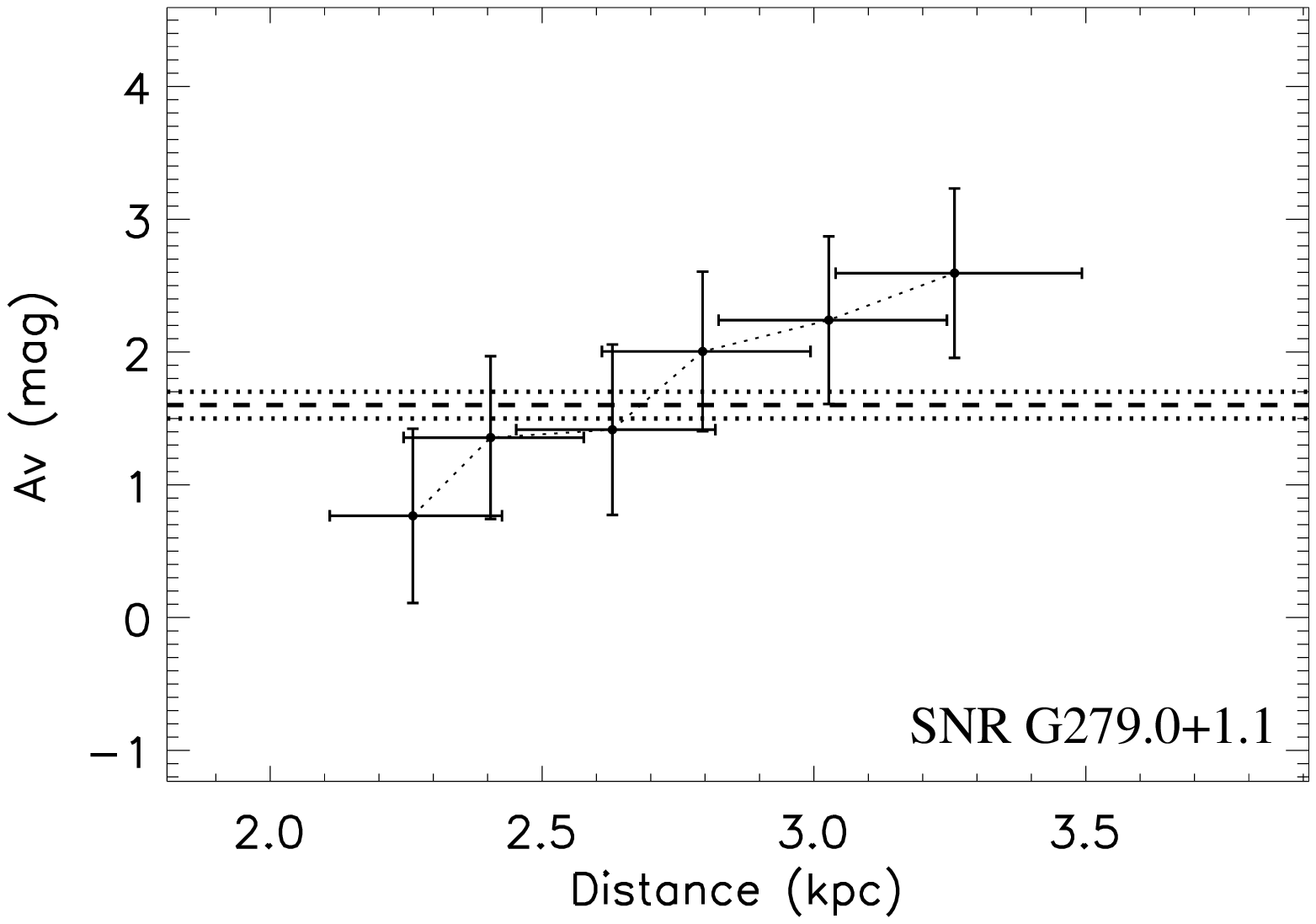}&
\vspace*{0.3cm}
\includegraphics[width = 0.5\textwidth]{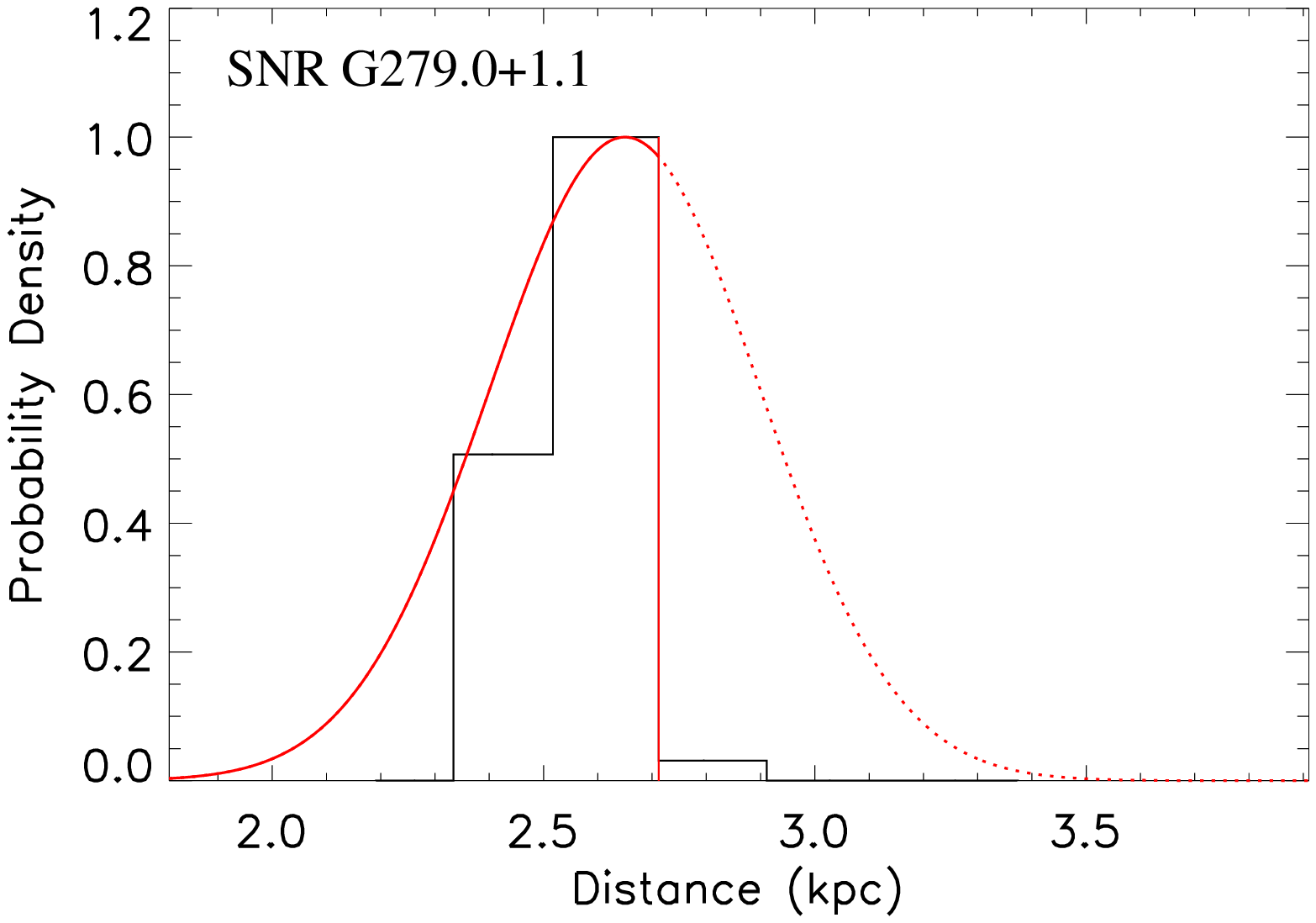}\\
\vspace*{0.3cm}
\includegraphics[width = 0.5\textwidth]{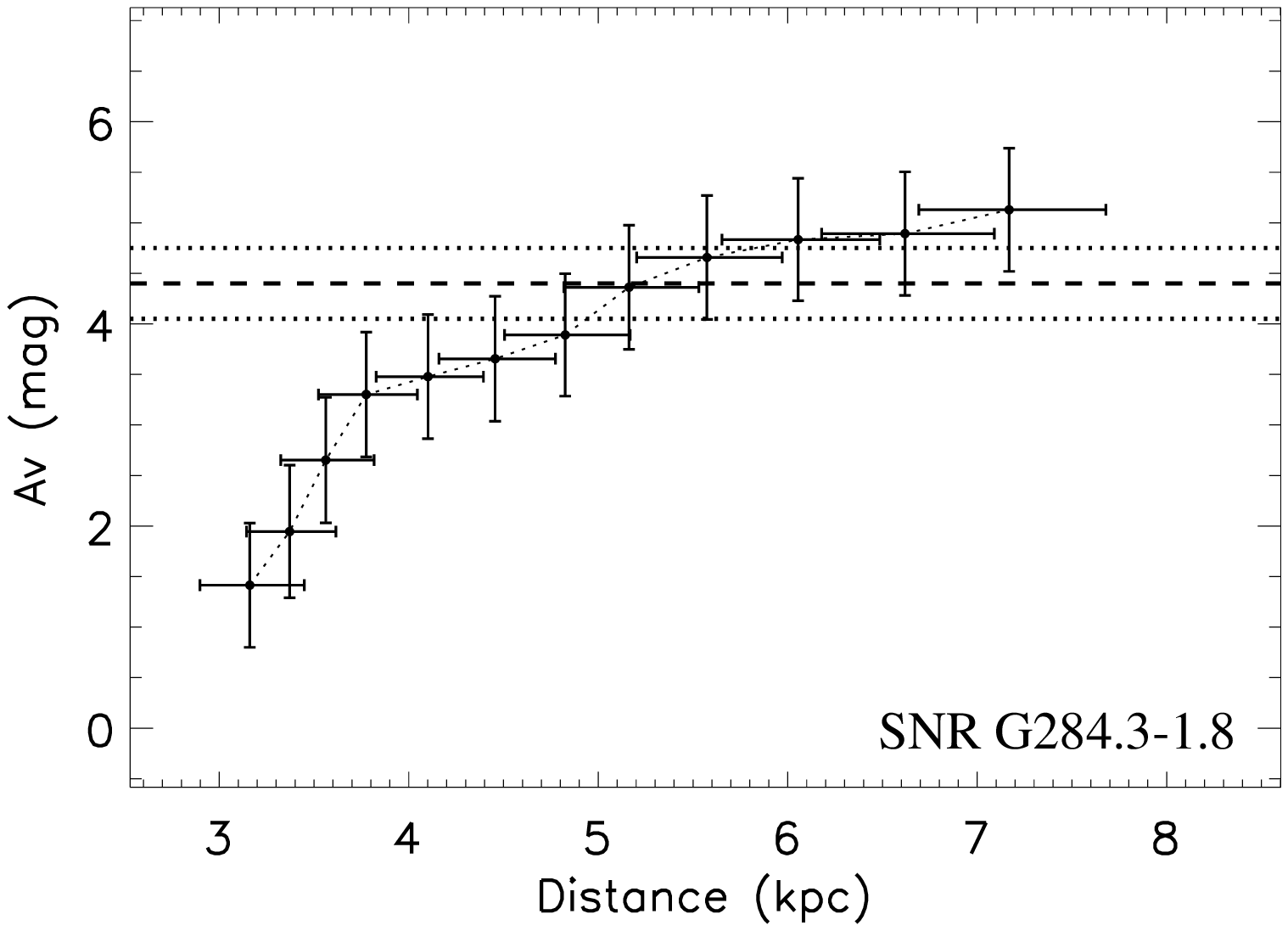}&
\vspace*{0.3cm}
\includegraphics[width = 0.5\textwidth]{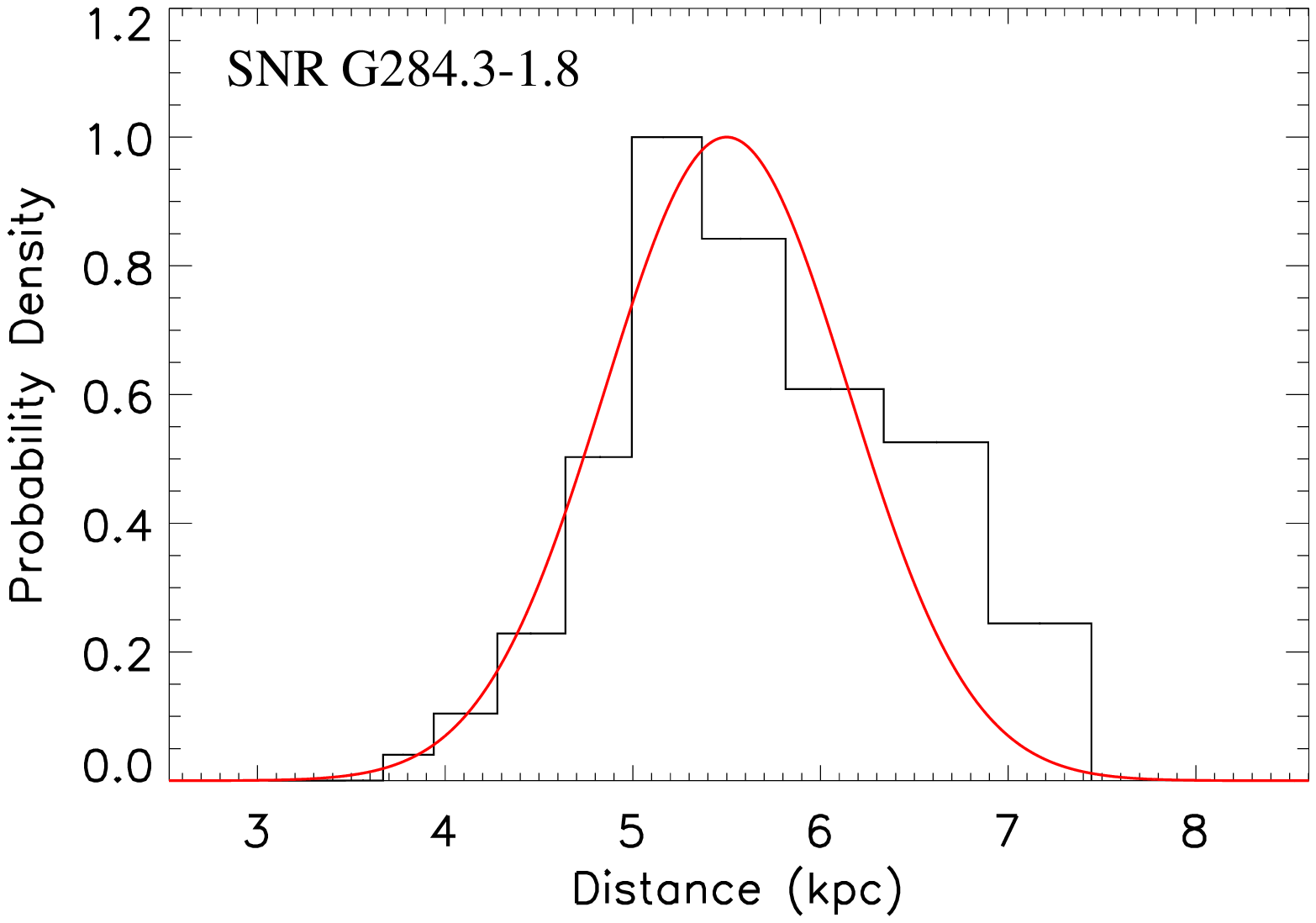}\\
\vspace*{0.3cm}
\includegraphics[width = 0.5\textwidth]{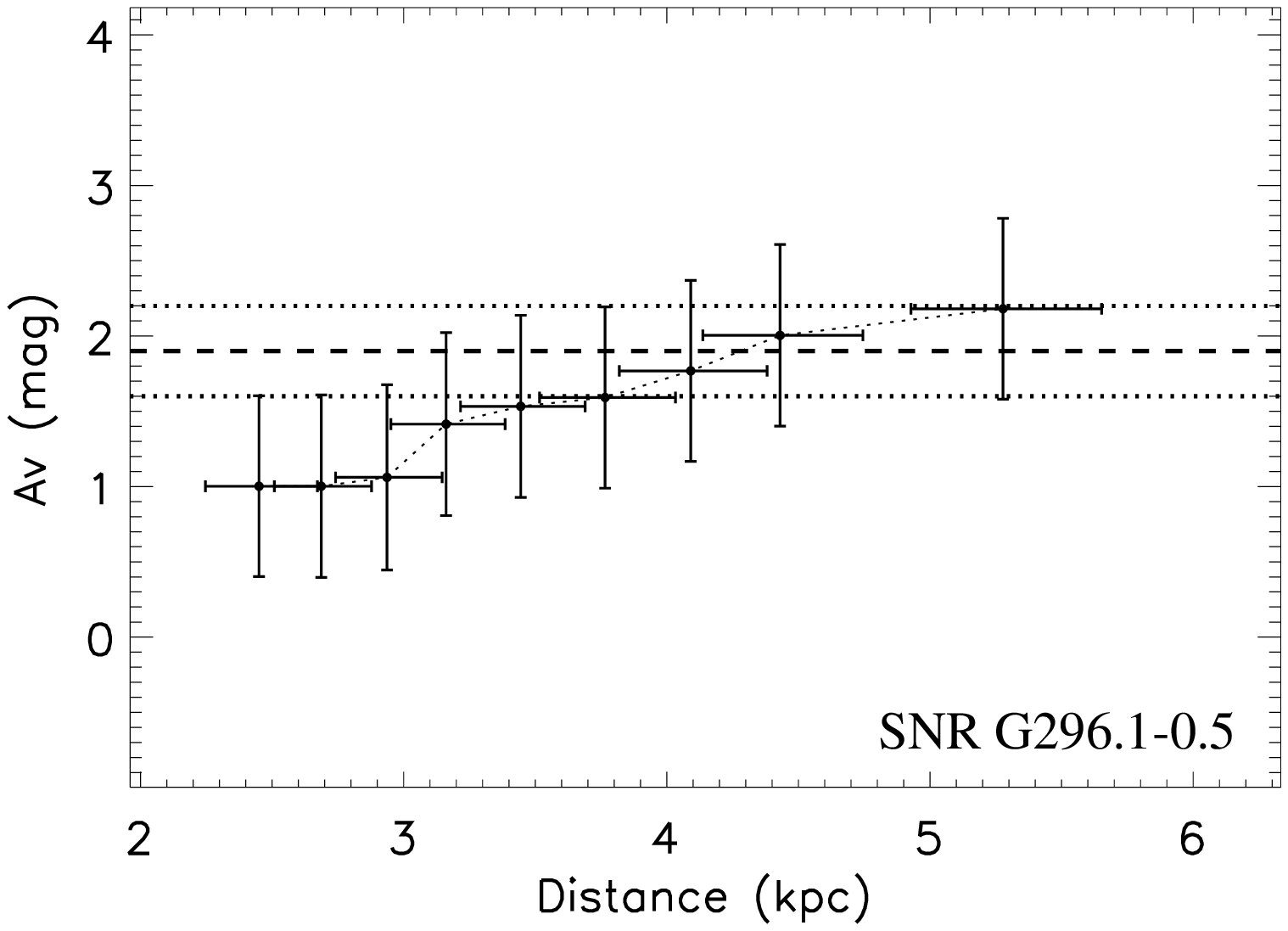}&
\vspace*{0.3cm}
\includegraphics[width = 0.5\textwidth]{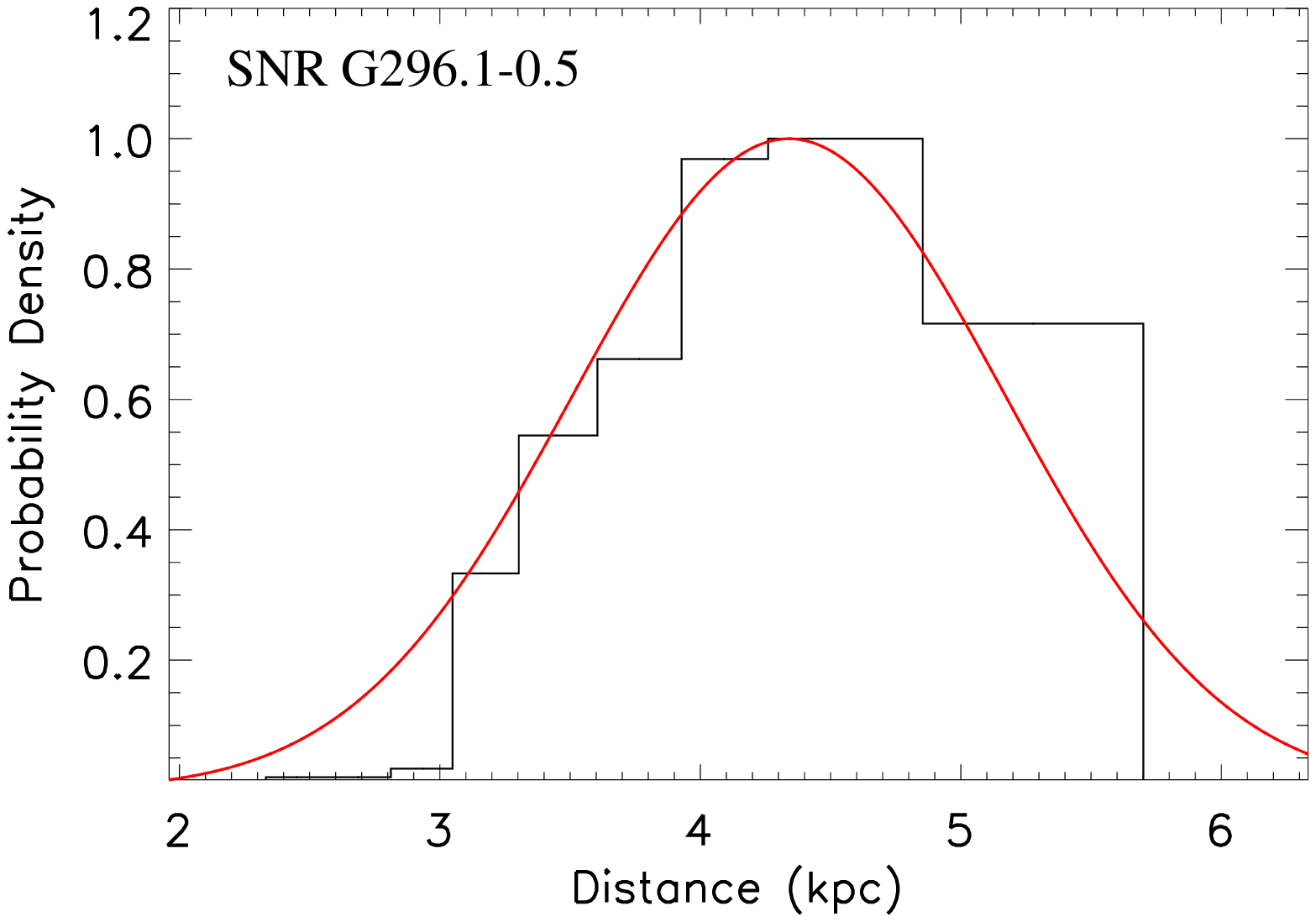}\\
\vspace*{0.3cm}
\includegraphics[width = 0.5\textwidth]{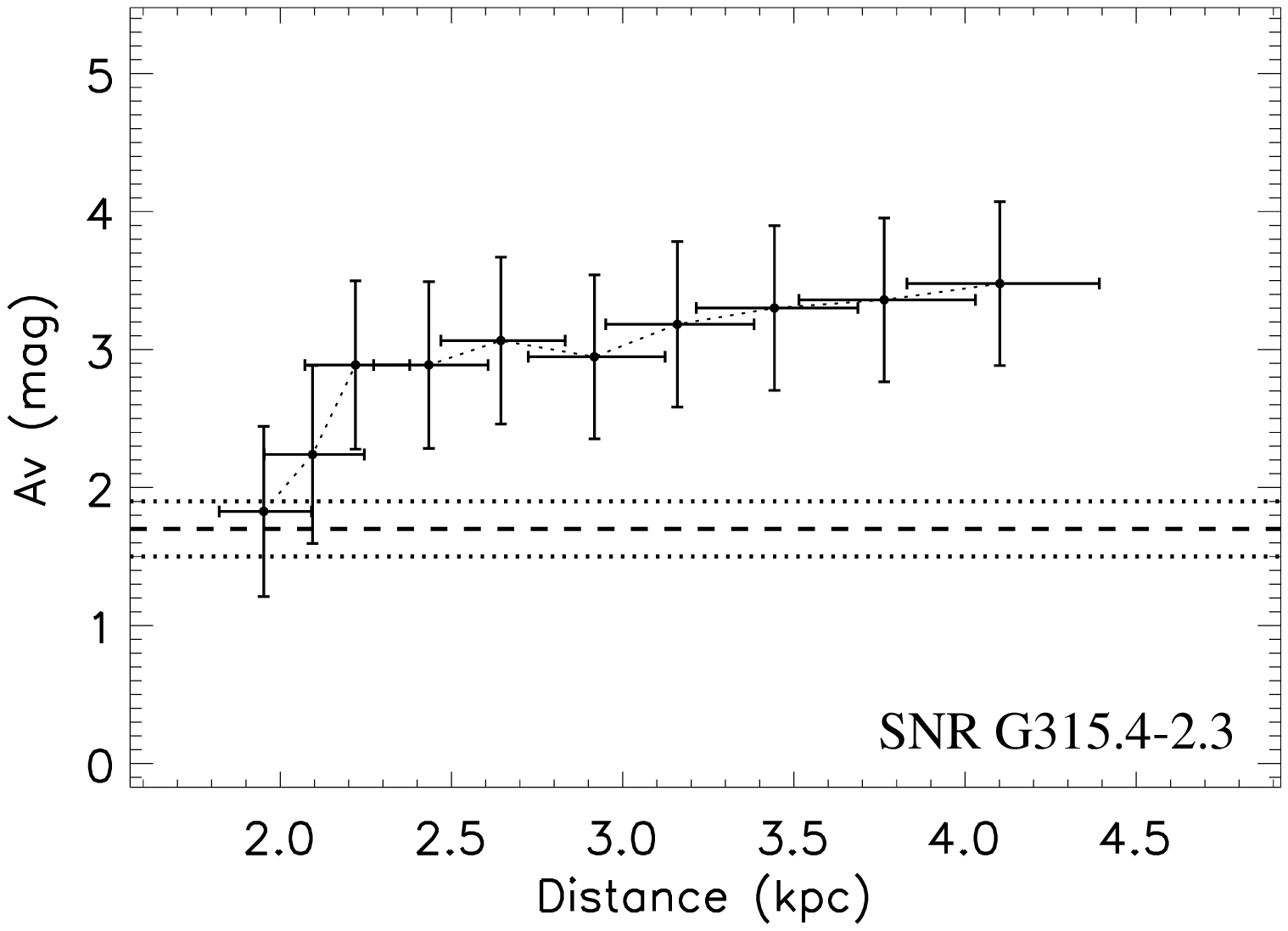}&
\vspace*{0.3cm}
\includegraphics[width = 0.5\textwidth]{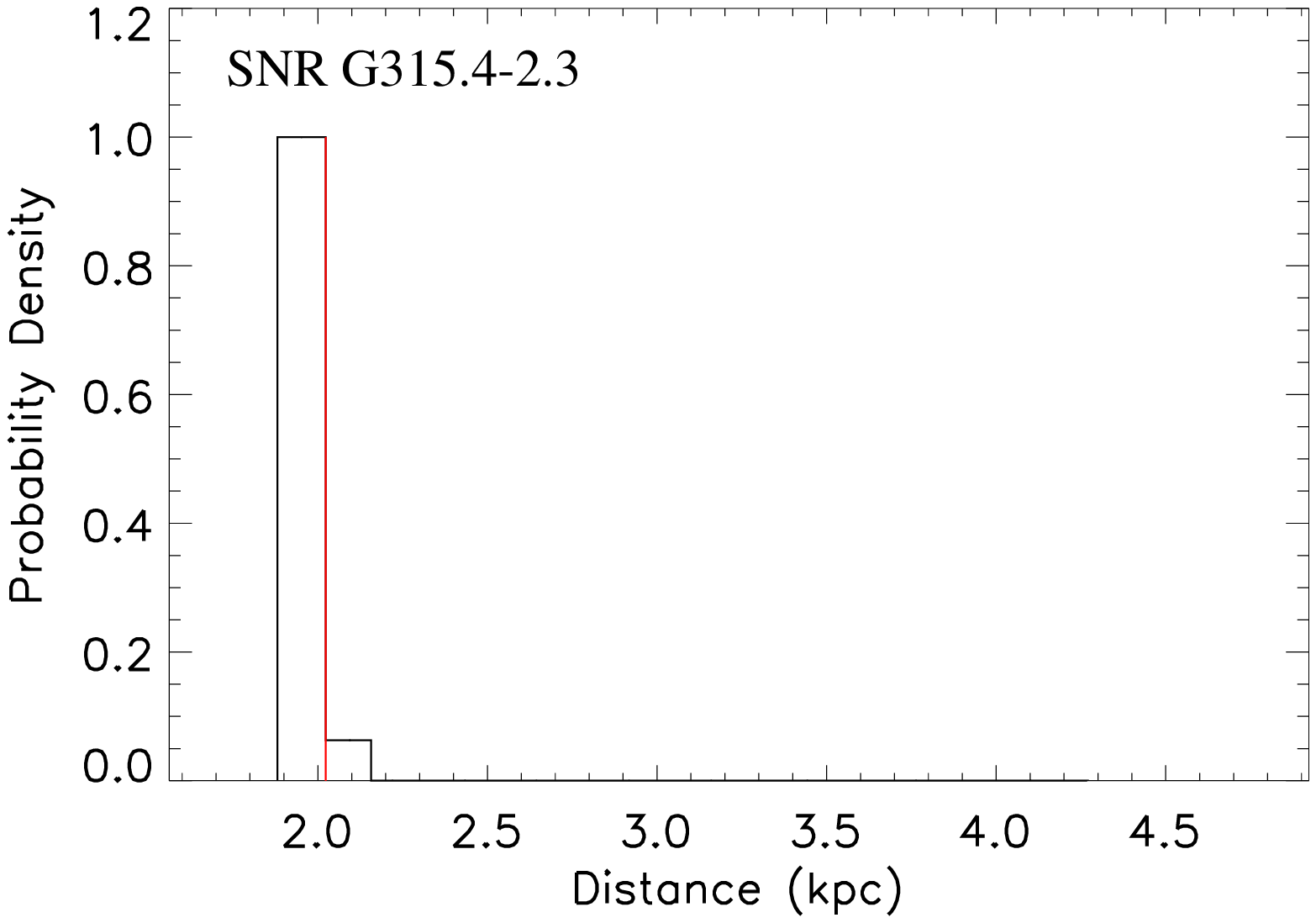}\\
\end{tabular}
\label{fig3}
\end{figure}
\addtocounter{figure}{-1}
\begin{figure}
\addtocounter{figure}{1}
\begin{tabular}{cc}
Fig. 3 continued&\vspace*{0.5cm}\\

\includegraphics[width = 0.5\textwidth]{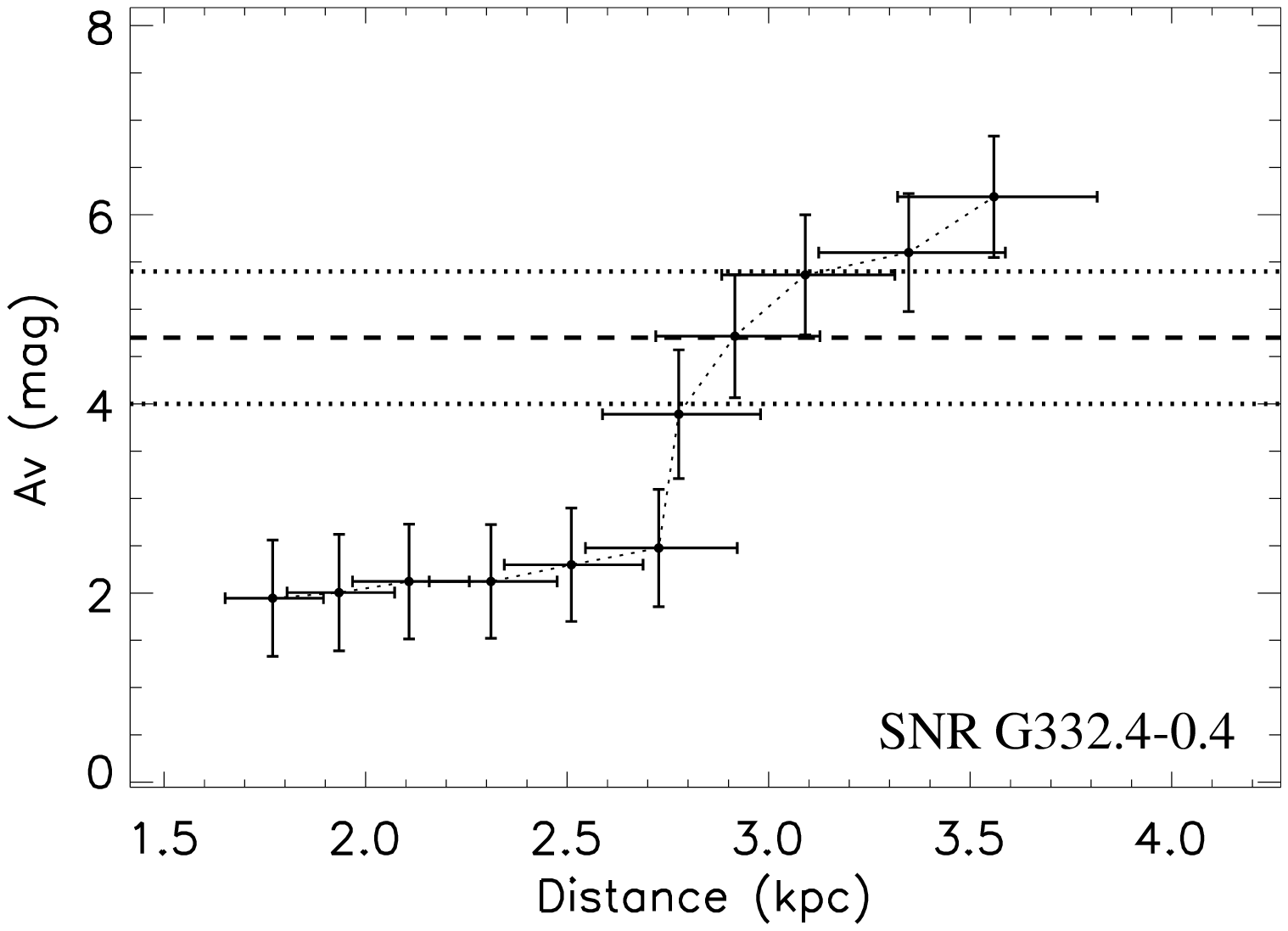}&
\vspace*{0.3cm}
\includegraphics[width = 0.5\textwidth]{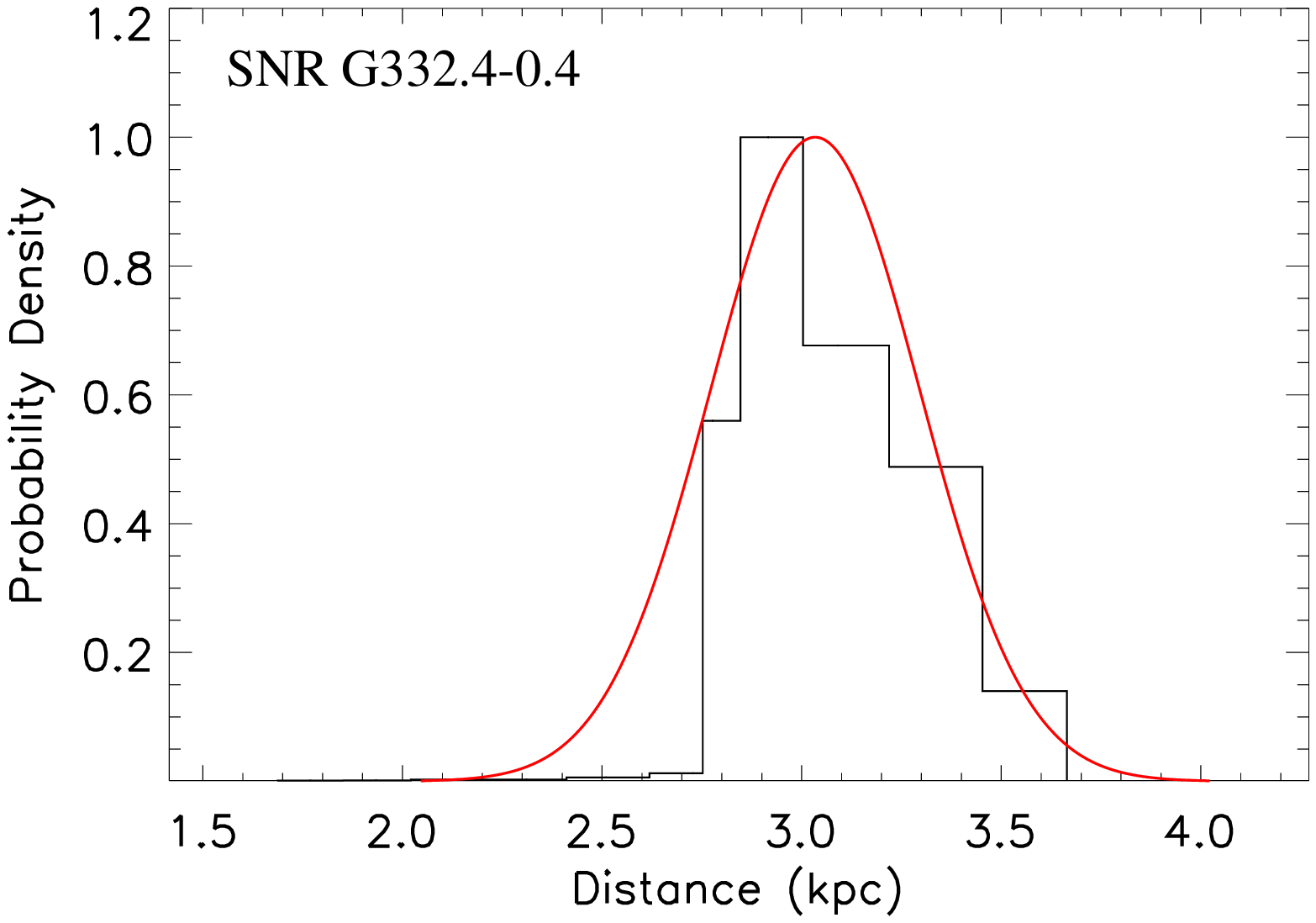}\\
\vspace*{0.3cm}
\includegraphics[width = 0.5\textwidth]{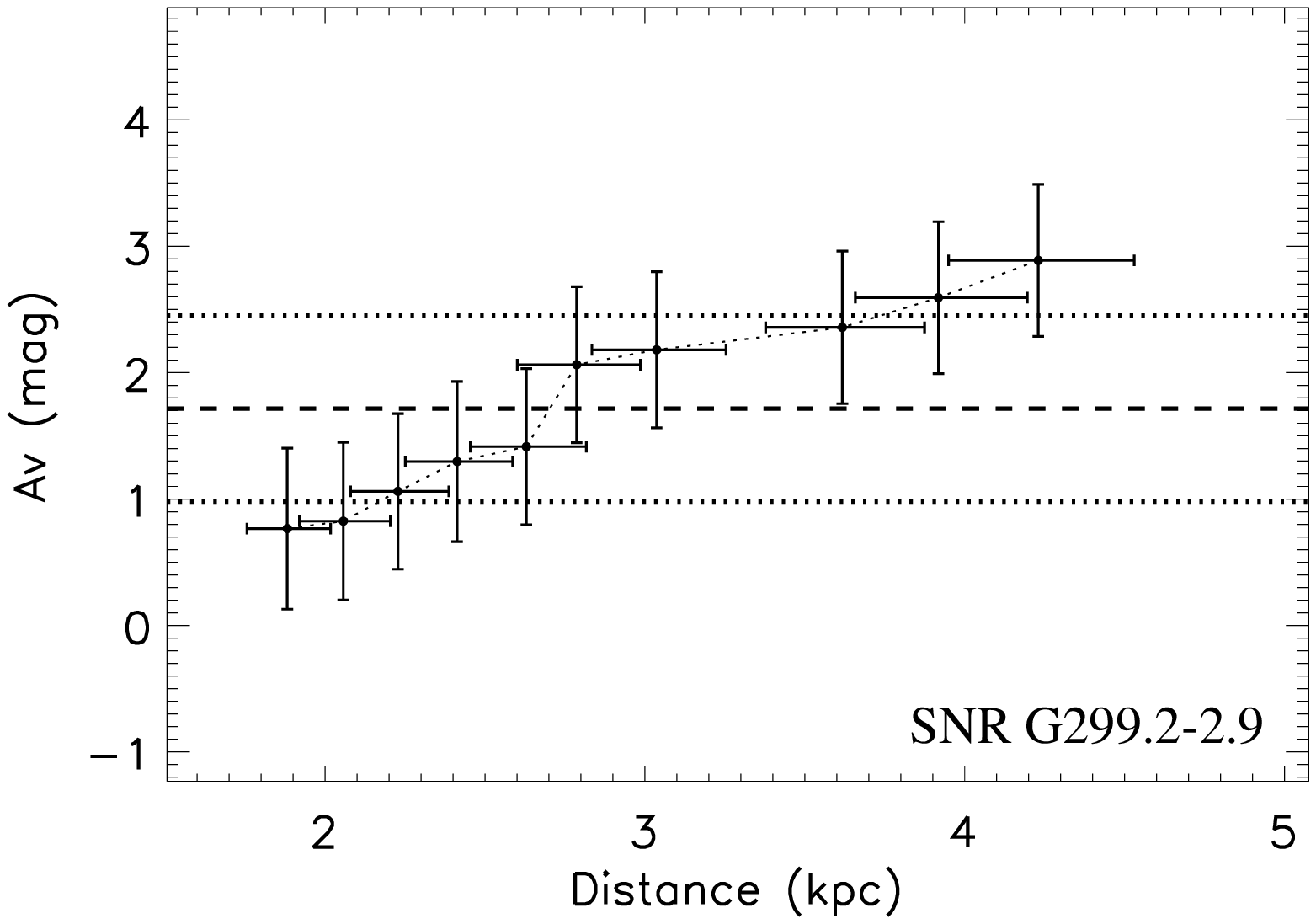}&
\vspace*{0.3cm}
\includegraphics[width = 0.5\textwidth]{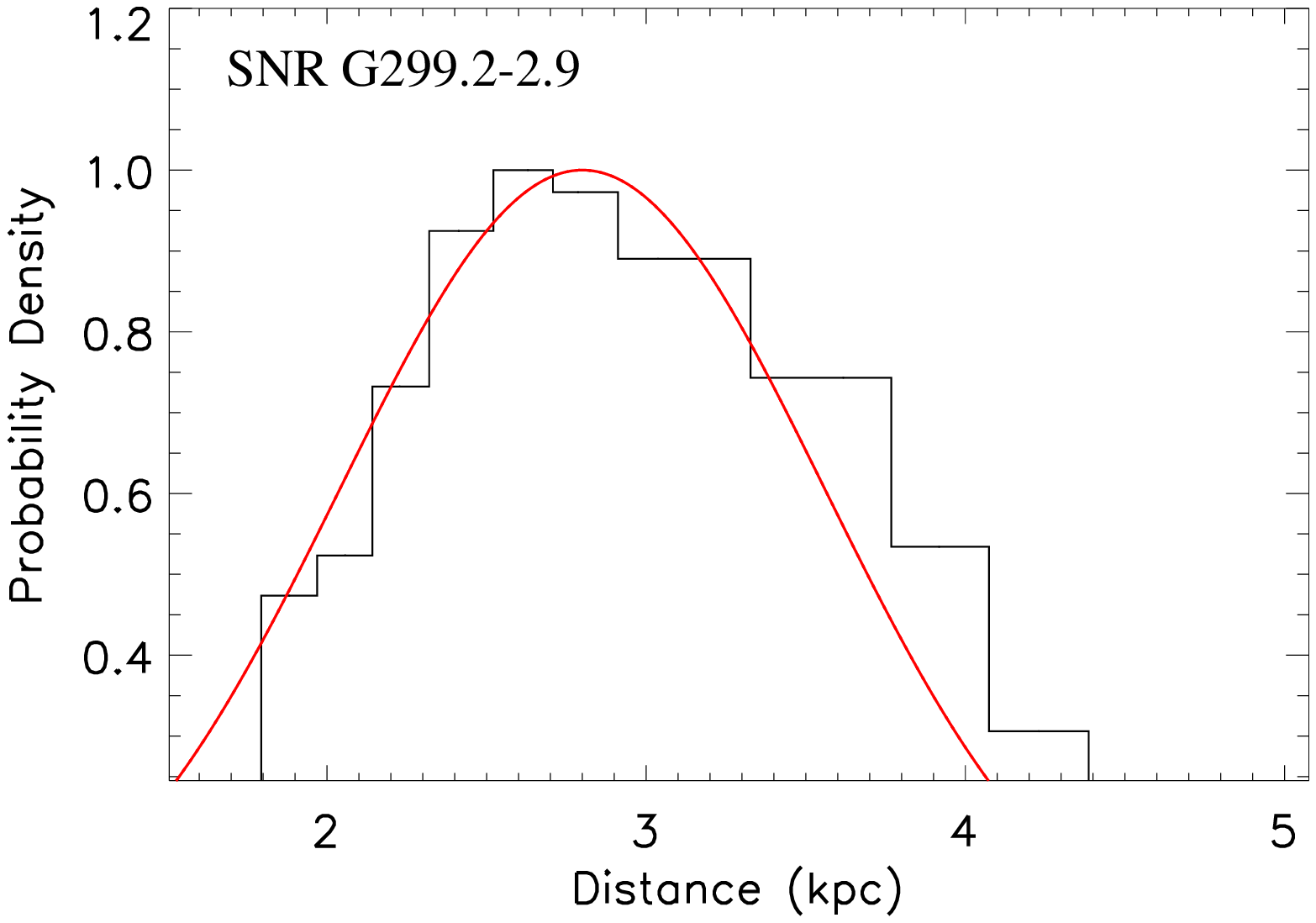}\\
\vspace*{0.3cm}
\includegraphics[width = 0.5\textwidth]{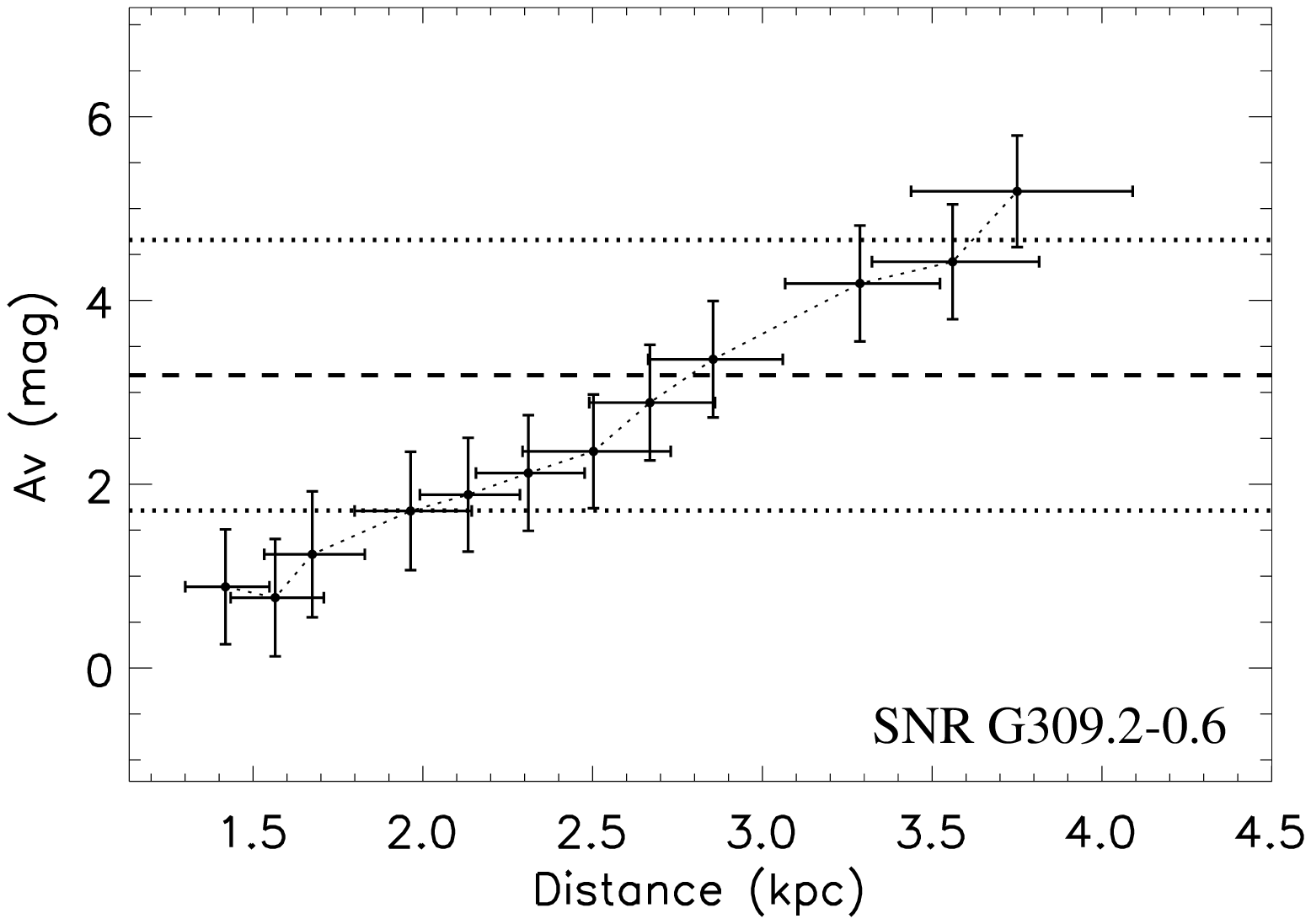}&
\vspace*{0.3cm}
\includegraphics[width = 0.5\textwidth]{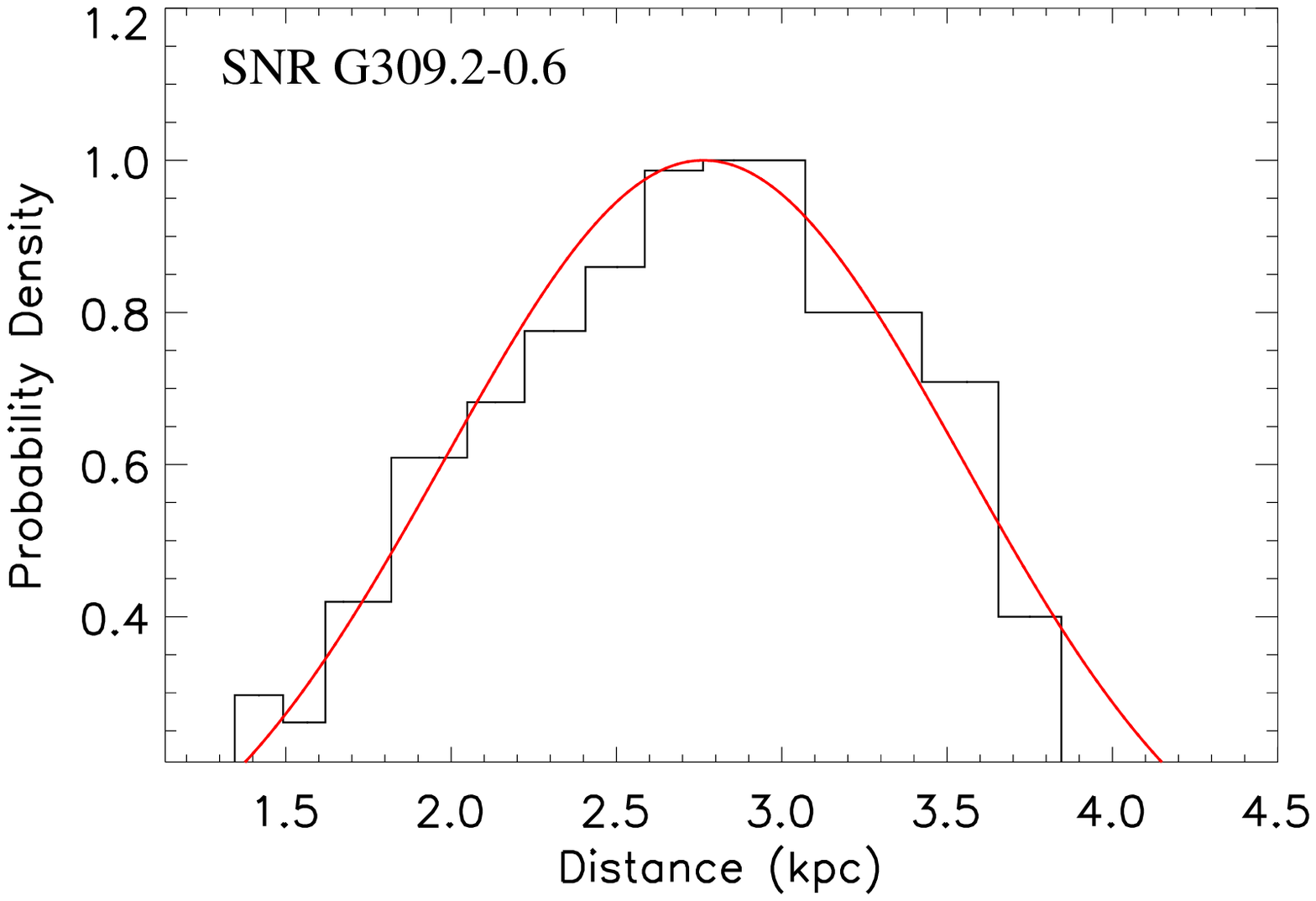}\\
\vspace*{0.3cm}
\includegraphics[width = 0.5\textwidth]{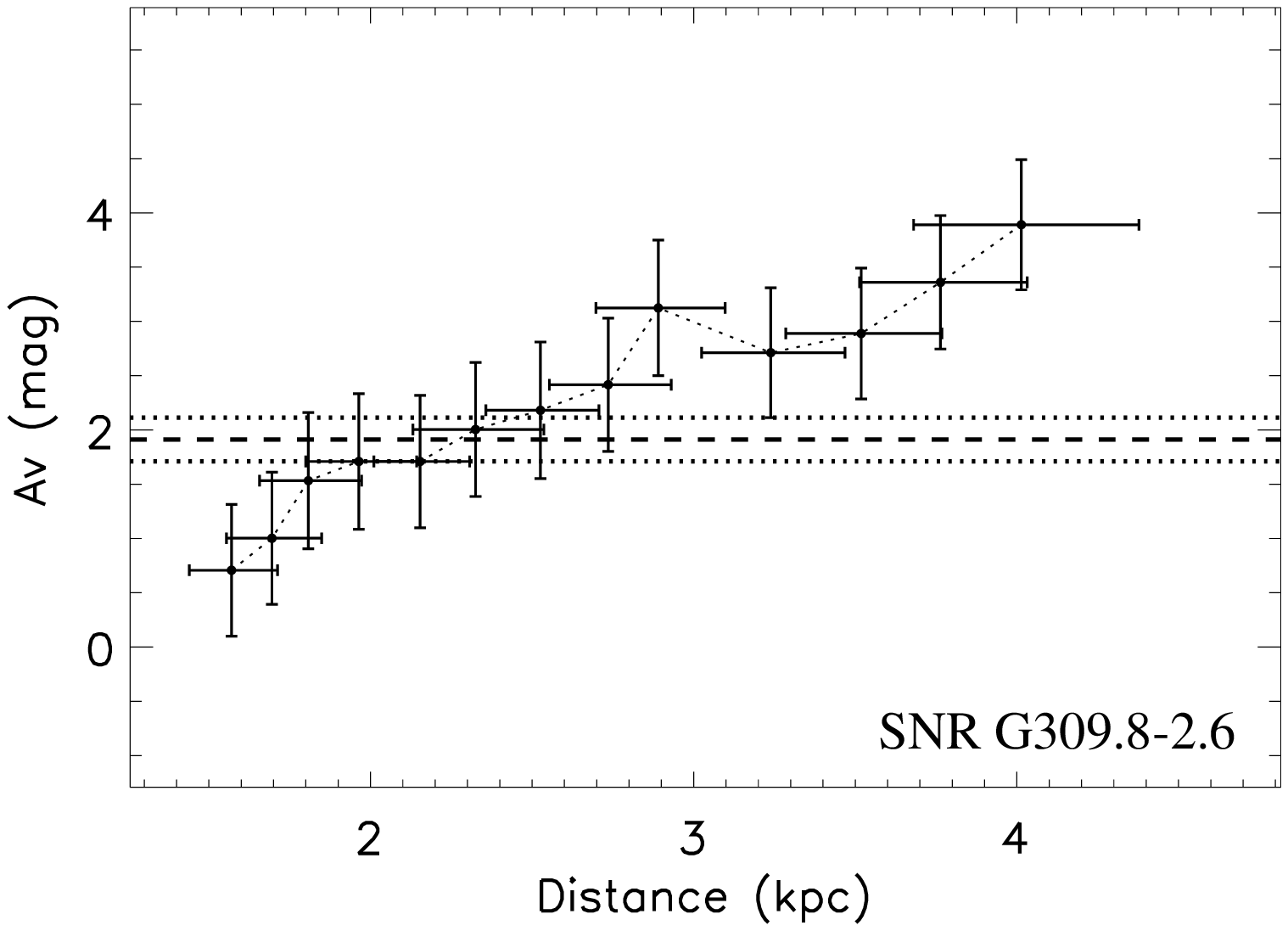}&
\vspace*{0.3cm}
\includegraphics[width = 0.5\textwidth]{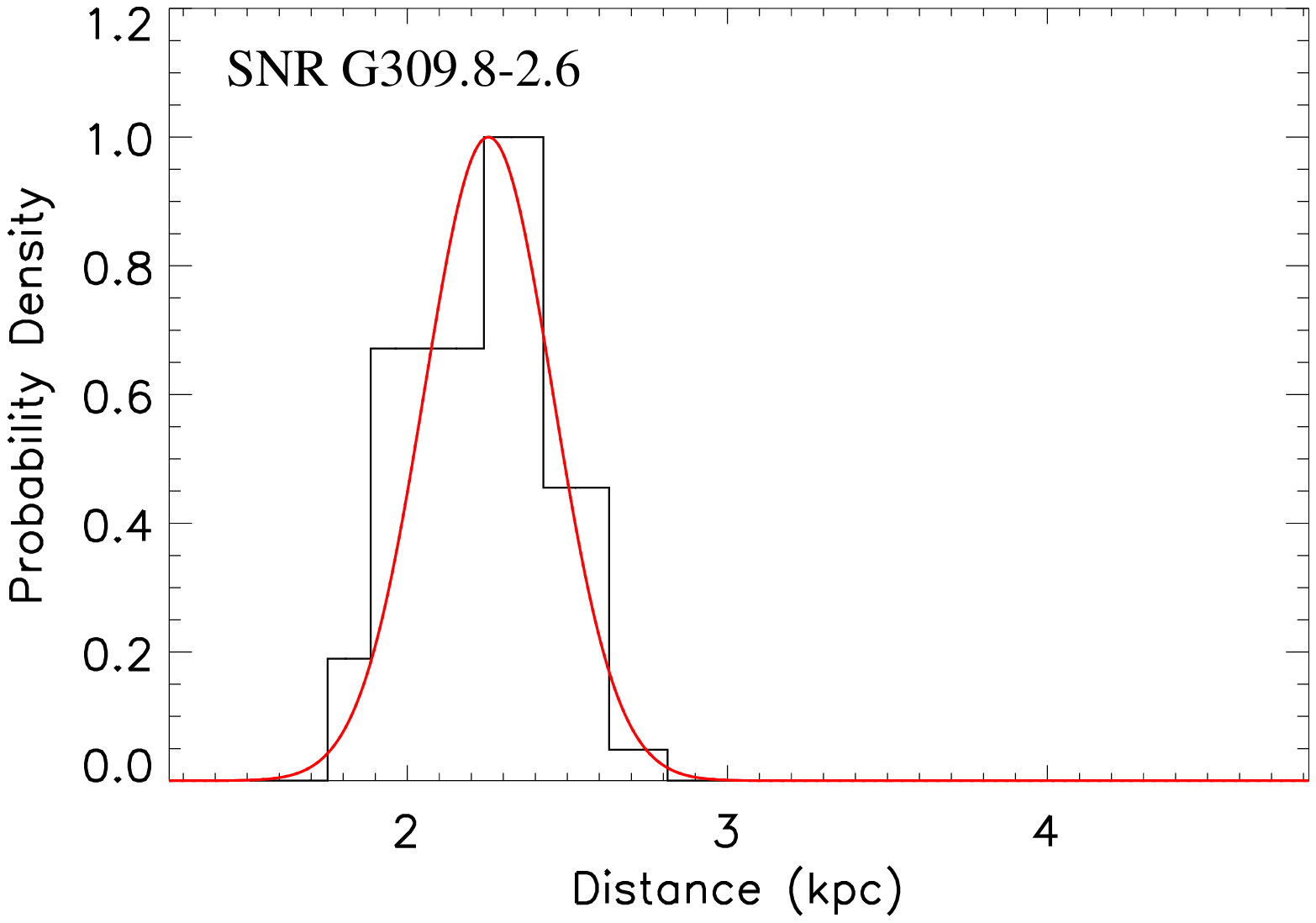}\\
\end{tabular}
\end{figure}

\textbf{G279.0+1.1}

The distance of G279.0+1.1 is estimated at 3 kpc by  blast wave method and $\Sigma$-D       
relation \citep{McKee1975,Stupar2009}. However, \citet{Stupar2009} warned that this distance should be treated with caution since \citet{McKee1975}'s estimate depends on the initial explosion energy of progenitor star and $\Sigma$-D relation might have an error up to 40\%. Our RC method estimates its distance as 2.7$\pm0.3$ kpc which coincides with the previous measurements.

\textbf{G284.3-1.8}

 G284.3-1.8 was suggested to associate with the pulsar PSR J1016-5857 (3 kpc， \citep{Kargaltsev2013}) or the $\gamma$-ray binary 1FGL J1018.6-5856 (5.6$_{-2.1}^{+4.6}$ kpc \citep{Napoli2011}).  We  estimate its distance at 5.5$\pm0.7$ kpc. The new distance  indicates that G284.3-1.8 is not likely to be associated with the PSR J1016-5857 at 3 kpc.
 
\textbf{G296.1-0.5}

G296.1-0.5 is a middle-aged SNR with complex structures. \cite{Longmore1977} estimated its distance between 3 and 5 kpc from  two independent ways of reddening  measurements and kinematic method. Distances calculated by
the $\Sigma$-D relation are 7.7, 6.6, 4.9 kpc from \citet{Caswell1983}, \citet{Case1998}, and \citet{Clark1973}, respectively. We show its distance of 4.3$\pm0.8$ kpc, which further supports the results of \citet{Longmore1977} and \citet{Clark1973}.

 \textbf{G315.4-2.3 (RCW 86)}
 
 \citet{Rosado1996} fitted the radial velocity of  H$_{\alpha}$ with  respect to the local standard of rest and converted it to a kinematic distance as 2.8$\pm0.4$ kpc for G315.4-2.3 based on Galactic rotation curve. \citet{Sollerman2003} followed this method with the data of high spectroscopic resolution and constrained the distance as 2.3$\pm0.2$ kpc. These distances were supported by the combination of the proper motion and the post-shock temperature \citep{Helder2013}. A smaller distance of $\rm 1.2_{-0.2}^{+0.2}$ kpc was determined with Sedov estimates \citep{Bocchino2000}. There is a cutoff in the probability distribution of distance because only the upper limit of $\rm A_V$ of G315.4-2.3 is  in the range of $\rm A_V$ traced by the RC stars. Therefore,the red line in the right panel for G315.4-2.3 is the upper distance limit, estimated to be  2.0 kpc by the RC method.
 
\textbf{G332.4-0.4 (RCW 103)}

The kinematic distance based on HI absorption spectrum is about 3.3 kpc for RCW 103 \citep{Caswell1975,Reynoso2004}. \citet{Ruiz1983}  estimated its distances around 6.5 kpc  by extinction measurement. However, the extinction traced by the RC stars infers  that this SNR locates at  3.0$\pm0.3$ kpc, which  is consistent with the kinematic distances.

 \textbf{G299.2-2.9}
 
We derive an RC distance of G299.2-2.9 at 2.8$\pm0.8$ kpc  for the first time.  A low  interstellar column density   in the line of sight hinted a very nearby distance ($\sim$0.5 kpc) based on the ROSAT Position Sensitive Proportional Counter observation \citep{Slane1996}. However, spatially resolved spectroscopy with the Chandra X-Ray Observatory shows a higher hydrogen column density ($\rm N_H \sim 3.5\times10^{21}cm^{-2}$), which suggested a distance of 2-11 kpc \citep{Park2007}. The RC distance is the best so far.

\textbf{G308.4-1.4}

\citet{Prinz2012} gave a detailed analysis of the distance  to G308.4-1.4: (1) the distance indicated by extinction is 5.9$\pm2.0$ kpc; (2) a kinematic distance is $2.0\pm0.6$ kpc or 12.5$\pm$0.7 kpc; (3) the Sedov-analysis-based distance is 9.8$^{+0.9}_{-0.7}$ kpc.  The RC distance of this SNR is 3.1$\pm0.3$ kpc, which is consistent with the low  limits of  extinction distance and kinematic distance.  

\textbf{G309.2-0.6} 

HI absorption against G309.2-0.6 yielded a kinematic distance from 5.4$\pm1.6$ to 14.1$\pm0.7$ kpc \citep{Gaensler1998}.  \citet{Rakowski2001} estimated the distance of this SNR to be $4\pm2$ kpc from the foreground hydrogen column density. The distance measured by the RC method is 2.8$\pm0.8$ kpc, which is consistent with the low limits of the  previous measurements.

\textbf{G309.8-2.6} 

 We estimate the distance of G309.8-2.6 at 2.3$\pm0.2$ kpc for the first time. The RC distance further supports that G309.8-2.6  is associated with  the very young pulsar PSR J1357-6429 located at 2.5 kpc \citep{Camilo2004}.
 
\begin{figure}
\centering
\includegraphics[angle=0,width=0.7\textwidth]{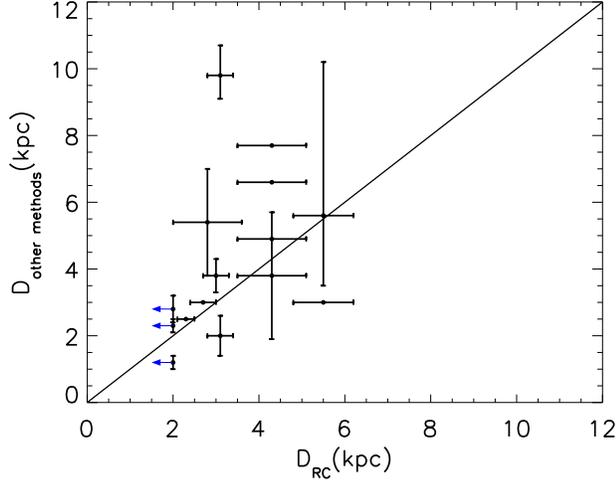}
\caption{ Comparison of the RC distances and the distances determined by other methods. The blue arrows represent the upper limit of distance for G315.4-2.3. The data is based on Table~\ref{tab1} and ~\ref{tab2}. }
\label{fig4}
\end{figure}

 We compare the  9 RC distances of SNRs with the distances from other methods  Fig.~\ref{fig4}. Note that there might be  several distance measurements  for one SNR and a few of them without uncertainty estimates.  The RC distances  are generally consistent with the previous measurements within the error bars. The distance uncertainties from the RC method  range from 10\% to 30\% and the accuracies of distances are significantly improved for 8 SNRs.
 
\normalem

\begin{acknowledgements}
We all acknowledge supports from  National Key R\&D Programs of China (2018YFA0404203) and NSFC programs (11603039, U1831128). We highly appreciate the anonymous referee for helpful comments.

\end{acknowledgements}
  
\bibliographystyle{raa}
\bibliography{bibtex}

\end{document}